\documentclass{aa}  

\usepackage{graphicx}
\usepackage{txfonts}
\usepackage[dvipsnames]{xcolor}
\usepackage{ulem}
\usepackage{amsmath}
\usepackage{upgreek}
\usepackage{tcolorbox}
\usepackage{hyperref}
\usepackage{verbatim}
\hypersetup{
    colorlinks=true,
    allcolors=blue
}
\newcommand{\GG}[1]{}

\newcommand{\vgp}{\Vec{\upmu}}

\newcommand{\eq}[1]{Eq.~\eqref{#1}}

\newcommand{\xib}{\eta}

\newcommand{\PI}{\hyperlink{paper1}{Paper I}}

\begin{document}

    \title{Properties of the ionisation glitch}
    \subtitle{II. Seismic signature of the structural perturbation}

   \author{Pierre S. Houdayer \thanks{\email{pierre.houdayer@obspm.fr}}
            \and Daniel R. Reese 
            \and Marie-Jo Goupil}

   \institute{LESIA, Observatoire de Paris, Université PSL, CNRS, Sorbonne Université, Université de Paris, 5 place Jules Janssen, 92195 Meudon, France}

   \date{Received 9 February 2022; accepted 13 May 2022}

% \abstract{}{}{}{}{} 
% 5 {} token are mandatory
 
  \abstract
  % context heading (optional)
  % {} leave it empty if necessary  
   {}
  % aims heading (mandatory)
   {In the present paper, we aim to constrain the properties of the ionisation region in a star from the oscillation frequency variation (a so-called glitch) caused by rapid structural variations in this very region. In particular, we seek to avoid the use of calibration based on stellar models thus providing a truly independent estimate of these properties. These include both the helium abundance and other physical quantities that can have a significant impact on the oscillation frequencies such as the electronic degeneracy parameter or the extent of the ionisation region.}
  % methods heading (mandatory)
   {Taking as a starting point our first paper, we applied structural perturbations of the ionisation zone to the wave equation for radial oscillations in an isentropic region. The resulting glitch model is thus able to exploit the information contained in the fast  frequency oscillation caused by the helium ionisation but also in the slow trend accompanying that of hydrogen. This information can directly be expressed in terms of parameters related respectively to the helium abundance, electronic degeneracy and extent of the ionisation region.}
  % results heading (mandatory)
   {Using a Bayesian inference, we show that a substantial recovery of the properties at the origin of the glitch is possible. A degeneracy between the helium abundance and the electronic degeneracy is found to exist, which particularly affects the helium estimate. Extending the method to cases where the glitch is subject to contamination (e.g. surface effects), we noted the importance of the slow glitch trend associated with hydrogen ionisation. We propose using a Gaussian process to disentangle the frequency glitch from surface effects.}
  % conclusions heading (optional), leave it empty if necessary 
   {}


   \maketitle

% ARTICLE PARTS
%-------------------------------------------------------------------
\section{Introduction \label{INTRO}}

When a star is subject to various forms of excitation, certain resonances may emerge which depend on its geometry and internal structure. When the star is spherically symmetric, each of these sustained oscillations can be described by quantum numbers such as its radial order $n$ and degree $\ell$ which reflect the oscillations' geometric properties. The overall way in which the pulsation frequencies $\nu_{n,\ell}$ vary with the radial order has been described with asymptotic developments \citep{Vandakurov1967,Tassoul1980}. These approximations have been very successful in describing solar-like oscillations, especially for large radial orders.

Besides this asymptotic trend, the advent of high-precision photometry with CoRoT \citep{Baglin2006}, \textit{Kepler} \citep{Gilliland2010,Lund2017} and now TESS \citep{Ricker2015,Stassun2019} has enabled the identification of more subtle contributions, including a particularly remarkable component, namely the ionisation glitch. Initially observed for the Sun, this deviation from the global trend manifests itself as an oscillatory component whose period, amplitude and damping reflects local properties of the ionisation region \citep{Gough1990,JCD1992,Gough2002}. Numerous studies have then exploited this particular signature in solar-like stars, in order to determine the position of their ionisation region \citep[eg.][]{Verma2014a,Broomhall2014,Verma2017} or the amount of helium it contains \citep[eg.][]{Vorontsov1992,PerezHernandez1994,Lopes1997,Verma2014a,Verma2019}. Most of the recent approaches are based on empirical models of the perturbation caused by the ionisation of helium \citep[e.g.][]{Monteiro2005,Houdek2007} and relate their parameters to actual physical quantities via calibration on realistic models \citep{Houdek2011,Verma2014a,Verma2019}. In our first article (\citealt{Houdayer2021}, hereafter \PI), we made various comments on this procedure and presented an alternative approach to model the ionisation region as well as the perturbation caused by a change in its properties. The derived structure is found to depend on three parameters that correspond respectively to the surface helium abundance, the electron degeneracy in the convective zone, and the extent of the ionisation region. Characterising the glitch based on such a model would thus allow us to extract physical quantities such as the helium abundance without the need for calibration, as well as other interesting constraints such as the electronic degeneracy in the star's convection zone.

In order to reach this objective, we establish, in the present paper, the relation between the properties of the ionisation region and the resulting glitch. The corresponding formalism (and the underlying assumptions) will be presented in Section~\ref{FREQ}, after having provided some details on the glitch modelling in Section~\ref{MOD}. Tests are subsequently carried out in order to assess how well the properties of the ionisation region may be recovered from a glitch. The results are described in Section~\ref{FIT} for an ideal case and Section~\ref{DISCUSSION} when a contamination is included. The closing Section~\ref{CONCLUSION} will be dedicated to our conclusion.

\section{Creating an ionisation glitch model \label{MOD}}

% MODEL SCHEME FIGURE --------------------------------------
   \begin{figure*}[!ht]
   \centering
   \includegraphics[width=15cm]{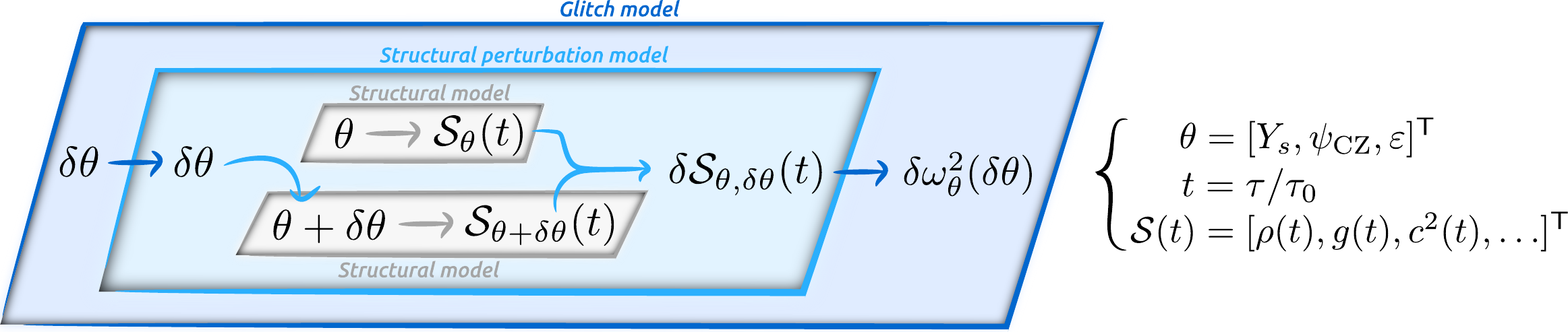}
   \caption{Schematic diagram of the method adopted here to derive a glitch model. The aim of Section~\ref{FREQ} is represented by the rightmost arrow, i.e. how to derive the glitch from a given structural difference.}
        \label{model_scheme.pdf}%
    \end{figure*}
% --------------------------------------------------------

\subsection{Modelling the ionisation region}

In \PI, we presented a thermodynamically derived model of the structure inside the ionisation region. This is done by means of an analytical approximation of the first adiabatic exponent $\Gamma_1(\rho,T,X_{i>1})$ inside an adiabatic region; the $X_{i>1}$ refer here to the abundances of all elements except hydrogen. When applied to a hydrogen-helium mixture (of respective abundances $1-Y$ and $Y$), solving the hydrostatic equilibrium as well as Poisson's equation defines entirely the associated structure (i.e., the density $\rho$, pressure $P$, sound speed $c, \ldots$) at any location $t$ (we will often refer to the position in the star by its normalised acoustic depth $0\leq t\leq 1$ in this paper). The resulting structural model, $\mathcal{S}(t)$, is therefore able to account for the impact of different helium amounts in the ionisation region. More specifically, this model is fully determined by three parameters: (1) $Y_s$, the surface helium abundance, (2) $\psi_\textrm{CZ}$, the value of the electron degeneracy parameter inside the convective zone and (3) $\varepsilon$, a small parameter defining the position and extent of the ionisation region. For a more compact notation, this parameterisation will hereafter be designated by the vector $\vec{\uptheta} = [Y_s,\psi_\textrm{CZ},\varepsilon]^\mathsf{T}$ (and thus the associated structural model $\mathcal{S}_{\vec{\uptheta}}(t)$).

In this paper, we proposed using a difference of these models $\delta\mathcal{S}_{\vec{\uptheta}, \delta\vec{\uptheta}}(t) = \mathcal{S}_{\vec{\uptheta}+\delta\vec{\uptheta}}(t) - \mathcal{S}_{\vec{\uptheta}}(t)$ instead of an ad hoc profile in order to model the structural perturbation causing the ionisation glitch. Indeed, in exchange for an increased numerical cost (i.e. the time needed to compute the model $\mathcal{S}_{\vec{\uptheta}+\delta\vec{\uptheta}}(t)$ for each $\delta\vec{\uptheta}$), this approach has many advantages such as: 
\begin{itemize}
    \item the possibility to account for more complex and realistic perturbations, 
    \item the ability to model the perturbation of any structural profile and not only that of the first adiabatic exponent $\Gamma_1$,
    \item the way it relates these perturbations to actual physical properties of the region,
    \item the fact that it is not limited to the impact of a pure helium change
\end{itemize} and this for a limited number of parameters (3). Beside the structural aspects raised here, we will see that the method also leads to benefits regarding the exploitation of glitches.

\subsection{Modelling the ionisation glitch using an ad hoc perturbation profile}

Many approaches to obtain an expression for the seismic glitch relies on formulae derived from the variational principle, which relate changes in oscillation frequencies to internal structure perturbations \citep{Dziembowski1990,Gough1991,Antia1994}. One then aims to eliminate certain contributions to obtain a final expression which depends only on perturbations that can be described analytically. Thus, a starting point to many derivations of theoretical forms of the frequency shift resides in neglecting the contribution of slowly varying perturbations like differences in pressure $\delta_xP/P$ or density $\delta_x\rho/\rho$ compared to $\delta_x\Gamma_1/\Gamma_1$ \citep[eg.][]{Monteiro2005,Houdek2007,Houdek2011} where $\delta_x$ designates a perturbation taken at fixed normalised radius $x = r/R$. As an example, the approximation used in Appendix A of \cite{Houdek2007} gives the resulting (quadratic) frequency difference with respect to a given reference:
\begin{equation}
    \label{eq:dnu2_HG07}
    \delta \omega^2 = \omega^2-\omega_\textrm{ref}^2 \simeq \frac{\displaystyle\int_V \delta_x\Gamma_1 P ~{\left(\mathrm{div~}\vec{\upxi}\right)}^2\,dV}{\displaystyle \int_V \rho ~\vec{\upxi}^\mathsf{T}\vec{\upxi}\, dV},
\end{equation} where $\omega = 2\pi\nu$ is the angular oscillation frequency, $V$ designates the volume of the star, and $\vec{\upxi}$ the Lagrangian displacement caused by the oscillation. Such an expression can be highly useful; indeed, having an analytical expression for the single perturbation involved $\delta_x\Gamma_1$ is sufficient to provide a parametric form of the glitch. If it is also possible to approximate the eigenfunction \citep[e.g. with the WKB formalism, cf.][]{Bender1978}, then a fully analytical expression for the frequency shift can be derived \citep{Monteiro1994,Monteiro1998,Houdek2007}, although the many assumptions needed to achieve this may affect the reliability of the approximation. The main drawback to this analytical development lies in its primary interest: it eliminates the need to use a complete model of the structure. From then on, it is no longer possible to fit frequency differences from a reference model ($\delta \nu$). Instead, the frequencies ($\nu$) must be fitted directly. Although the frequency shift can be approximated by Eq.~\eqref{eq:dnu2_HG07}, the reference frequencies, $\nu_\textrm{ref}$, are not explicitly defined in this case. Rather than trying to model them roughly (e.g. via an asymptotic relationship), it is preferable to consider second differences in order to limit the error introduced \citep{Gough1990}: 
\begin{equation}
    \label{eq:d2nu}
    d^2\nu_{n,\ell} = \nu_{n+1,\ell}-2\nu_{n,\ell}+\nu_{n-1,\ell}.
\end{equation}

Applied to Eq.~\eqref{eq:dnu2_HG07}, the use of the second differences leads to
\begin{equation}
    \label{eq:d2nu_HG07}
    d^2\nu \simeq {\left(\Delta\nu\right)}^2\frac{d^2\mathcal{F}}{d\nu^2}+d^2\nu_\textrm{ref},
\end{equation} where $\Delta\nu$ corresponds to the asymptotic large separation and $\mathcal{F}(\nu)$ an analytical approximation of 
\begin{equation}
    \label{eq:Fnu}
    \mathcal{F}(\nu) \simeq \delta\nu \simeq \frac{\displaystyle\int_V \delta_x\Gamma_1 P ~{\left(\mathrm{div~}\vec{\upxi}\right)}^2\,dV}{8\pi^2\nu \displaystyle \int_V \rho ~\vec{\upxi}^\mathsf{T}\vec{\upxi} \, dV}.
\end{equation}

This way, the $d^2\nu_\textrm{ref}$ component is generally assumed to be \textit{smooth}, in the sense that it can be the result of a second derivative applied to the oscillation frequencies of a ``glitch-free'' model. It is thus usually eliminated from $d^2\nu$ by fitting a slowly varying function. Note that a strong assumption made here is the persistence of the linear perturbation regime, and thus that the perturbed model differs only slightly from the glitch-free model.\newline

\subsection{Modelling the ionisation glitch using a structural perturbation model}

In contrast, the structural perturbation model $\delta\mathcal{S}_{\vec{\uptheta}, \delta\vec{\uptheta}}(t)$ on which we rely allows us to express the perturbation of each structural quantity, so that we are no longer limited to a formula that depends only on $\delta_x\Gamma_1$. It is thus possible to rely on a more comprehensive form than Eq.~\eqref{eq:dnu2_HG07}, in which some of the contributions to the glitch were neglected. Also, it is possible to numerically compute the reference frequencies $\omega_\textrm{ref}$ as well as the eigenfunctions $\vec{\upxi}$ based on $\mathcal{S}_{\vec{\uptheta}}(t)$. With this in mind, we will try to construct a glitch model that makes the link explicit between a change in physical properties, $\delta\vec{\uptheta}$, and frequency perturbations $\delta\omega_{\vec{\uptheta}}^2(\delta\vec{\uptheta})$. The link between $\delta\vec{\uptheta}$ and the structural perturbation $\delta\mathcal{S}_{\vec{\uptheta}, \delta\vec{\uptheta}}(t)$ having been established in our first paper, it is now a matter of deriving an expression analogous to Eq.~\eqref{eq:dnu2_HG07} in order to express the relation between this structural perturbation and the glitch. A schematic representation of our approach is shown in Fig.~\ref{model_scheme.pdf}.

\section{Impact of structural perturbations on oscillation frequencies \label{FREQ}}

We aim to describe here the behaviour of the oscillation frequencies when disturbed by a perturbation of the equilibrium quantities in the ionisation region. We will rely on the following two assumptions:  
\begin{enumerate}
    \item The perturbation is localised in a convective zone, meaning that we will consider the region to be strictly isentropic.
    \item The perturbation is close enough to the surface so that we can assume the displacement (at least for low degree pulsation modes) to be purely radial.
\end{enumerate}

These approximations lead to the following simplifications in the calculations:
\begin{align}
    \ell ~&=~ 0 \\
    \gamma ~&\equiv~ \frac{d\ln P}{d\ln \rho} = \left(\frac{\partial\ln P}{\partial\ln \rho}\right)_S \equiv \Gamma_1
\end{align}and therefore the exact cancellation of both Lamb $S_\ell$ and Brunt-Väisälä $\mathcal{N}$ frequencies. Within this framework, the remaining constraints on the wave propagation (which are needed in order to define a cavity for a standing wave) will be grouped under the term cut-off frequency and denoted $\omega_c$.

\subsection{The wave equation \label{WAVE}}

Under the assumption of adiabatic and radial pulsations, the wave equation relating the oscillation frequency $\omega$ to the displacement amplitude $\xi$ can be reduced to a Schrödinger equation expressed on the acoustic depth $\tau \equiv \int_r^R dr'/c$  \citep{Roxburgh1994a,Roxburgh1994b}:
\begin{equation}
    \label{eq:wave3}
    \frac{\partial^{2} \lambda\xi}{\partial \tau^{2}}+\left(\omega^{2}-\omega_c^2\right)\lambda\xi=0,
\end{equation}with, 
\begin{equation}
    \label{eq:wcdef_r}
    \begin{split} 
        \omega_c^2 = \frac{1}{4}\left(\frac{2c}{r}-\frac{g}{c}-\frac{d c}{d r}\right)^{2}-\frac{c}{2} \frac{ d }{ d r}\left[\frac{2c}{r}-\frac{g}{c}-\frac{d c}{d r}\right] -4 \pi G \rho 
    \end{split}
\end{equation}and the rescaling variable $\lambda$ defined as:
\begin{equation}
    \label{eq:lambdadef}
    \lambda^2 \equiv -\frac{dm}{d\tau} = 4\pi r^2\rho c.
\end{equation}

In order to relate all the quantities that are homogeneous to a frequency in equation \eqref{eq:wcdef_r}, it is useful to introduce
\begin{equation}
    \label{eq:freqscale}
	F_s \equiv \frac{d\ln |s|}{d\tau},
\end{equation} the \textit{frequency scale} of any structural quantity $s$. In this formalism, the cut-off frequency can be written as a quadratic form over few frequency scales :
\begin{equation}
    \label{eq:wcdef}
    \omega_c^2 = 2{F_r}^2+\frac{{F_\rho}^2}{4}+\frac{{F_c}^2}{4}+3F_rF_\rho-2F_rF_c+\frac{F_\rho F_g}{2} +\frac{F_cF_{dc/dr}}{2}
\end{equation}by making use of the following properties : 
\begin{align*}
    F_{s_1s_2} ~&=~ F_{s_1}+F_{s_2}, \\
    F_r ~&=~ -c/r \quad \mathrm{and} \quad \frac{dF_r}{d\tau} ~=~ F_r\,(F_c-F_r),\\
    F_\rho ~&=~ g/c ~~\quad \mathrm{and} \quad \frac{dF_\rho}{d\tau} ~=~ F_\rho\,(F_g-F_c), \\
    F_c ~&=~ -\frac{dc}{dr} \!\;\quad \mathrm{and} \quad \frac{dF_c}{d\tau} ~=~ F_c\,F_{dc/dr}.
\end{align*}

% WC2 FIGURE --------------------------------------
   \begin{figure}
   \centering
   \includegraphics[width=9cm]{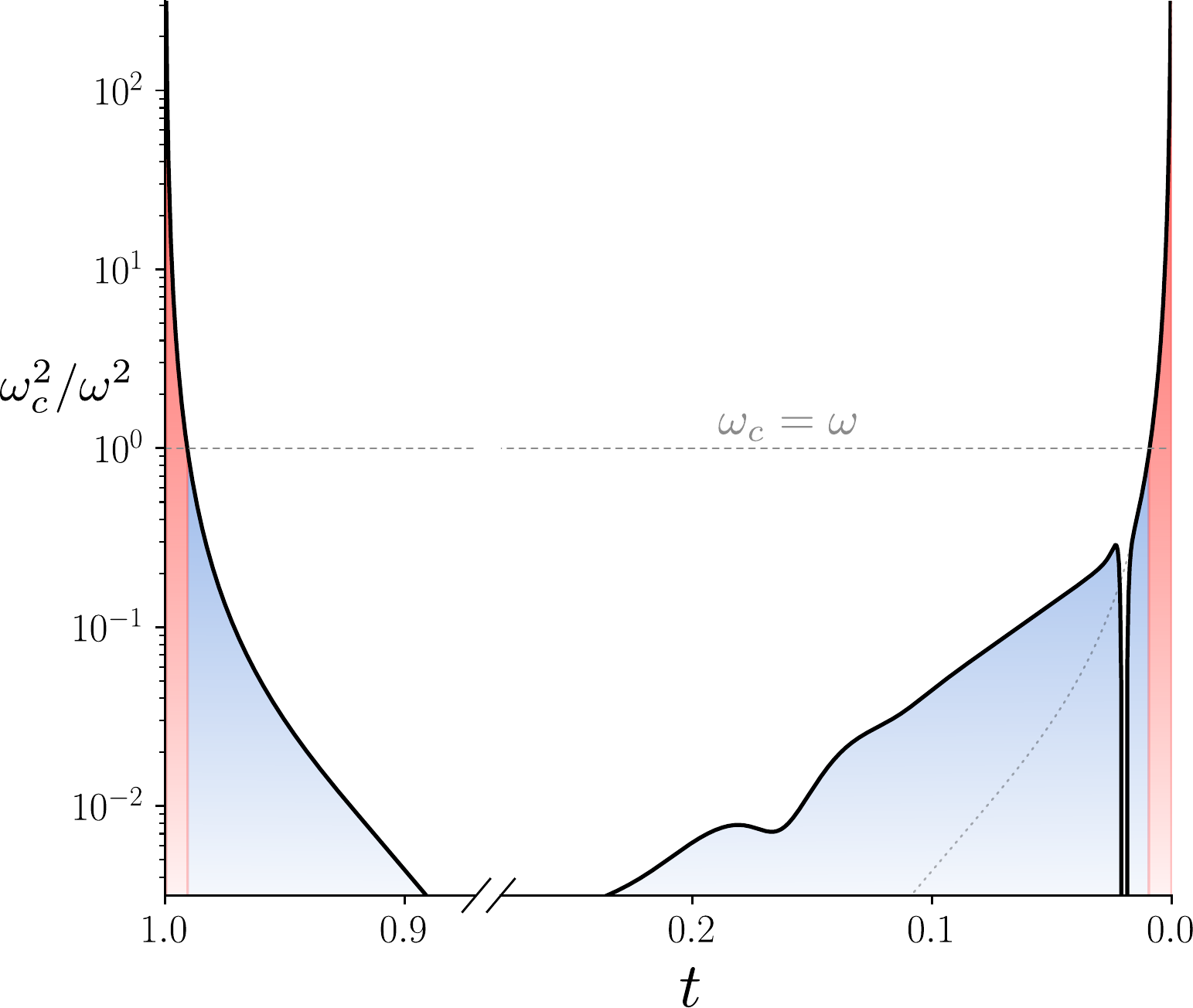}
   \caption{Example of the ratio $\omega_c^2/\omega^2$ in a solar-like model with respect to the normalised variable $t = \tau/\tau_0$ (the surface is located on the right-hand side and the centre is made visible by a cut on the axis). In the above plot, we have chosen a typical solar pulsation frequency $\omega = 2\pi\times 3000~\mu$Hz. The wave propagates in the blue shaded region ($0.01 < t < 0.99$) below the horizontal dashed line ($\omega_c = \omega$) while being damped in the red shaded area. The dotted line near the surface has been added for the sake of comparison and represents the symmetric counterpart to the curve near the centre.  It overlaps with the actual curve down to the ionisation region.}
        \label{omegac2.pdf}%
    \end{figure}
% --------------------------------------------------------

From Eq. \eqref{eq:wcdef}, we observe that the constraints to the wave propagation can be both of a geometrical (involving the terms in $F_r, F_g \sim c/r$ which dominate near the star's centre) and structural nature (composed of the terms in $F_\rho$, $F_c$ or $F_{dc/dr} \sim g/c$, dominating in the upper layers). A representation of the ratio between the resulting cut-off and oscillation frequencies $\omega_c^2/\omega^2$ with respect to the normalised acoustic depth $t = \tau/\tau_0$ ($\tau_0 = \int_0^Rdr/c$ being the star's acoustic radius) is given in Fig.~\ref{omegac2.pdf}. From \eq{eq:wave3}, we see that this ratio contains all the information about the propagation: as long as its value remains below $1$ (blue area), the wave will adopt an oscillatory behaviour whereas it will be evanescent if it exceeds this limit (red area). The variations close to the surface, which break the symmetry between the star's centre and surface, constitute a typical signature of the ionisation region.

Before going further, let us keep in mind that we are interested here in describing the behaviour of the glitch and not directly of the frequencies that appear in \eq{eq:wave3}. To this end, it is more convenient to normalise this relation by introducing quantities that can be compared from one structure to another. Accordingly, we rewrite the above Eq.~\eqref{eq:wave3} in terms of the following normalised variables: $\sigma  = \omega\tau_0$, $\sigma_{c} = \omega_c\tau_0$, $t = \tau/\tau_0$ and $\xib = \lambda\xi/\|\lambda\xi\|$ with the notation:
\begin{equation}
    \|\lambda\xi\| = \sqrt{\displaystyle\frac{1}{\tau_0} \int_0^{\tau_0} {(\lambda\xi)}^2\,d\tau}.
\end{equation}

The wave equation verified by the normalised eigenfunction $\xib(t,\sigma)$ can now be rewritten as follows:
\begin{equation}
    \label{eq:wave_ad}
    \frac{\partial^{2} \xib}{\partial t^{2}}+\left(\sigma^2-\sigma_c^2\right)\xib=0.
\end{equation}

\subsection{Obtaining a glitch model}

Taking the product of \eq{eq:wave_ad} and $\xib$, and integrating the result from $t = 0$ to $1$, we can naively write (after integrating by parts):
\begin{equation}
    \label{eq:intwave_ad0}
    \sigma^2_\mathrm{mod} = \int_0^1 \left(\sigma_c^2\,\xib^2+{\left(\frac{\partial\xib}{\partial t}\right)}^2\right)dt-\left[\xib\frac{\partial\xib}{\partial t}\right]_0^1,
\end{equation}remembering that $\xib$ is normalised and therefore $\int_0^1\xib^2\,dt = 1$. It is possible to show that the part in brackets can be considered to be null as long as it is possible to find a thin layer in both the surface and centre where the variations of $\Gamma_1$ are negligible. Since this layer can be as thin as desired, we will consider hereafter the following approximation:
\begin{equation}
    \label{eq:intwave_ad}
    \sigma^2_\mathrm{mod} = \int_0^1\left(\sigma_c^2\,\xib^2+{\left(\frac{\partial\xib}{\partial t}\right)}^2\right)dt.
\end{equation}

It is clear that the actual oscillation frequencies ($\sigma$) can significantly differ from the modelled expression of oscillation frequencies ($\sigma_\mathrm{mod}$) provided above. Indeed, if the assumptions of isentropy and radial oscillations seem coherent for pulsations located in a surface convective zone (and a fortiori in the ionisation region), these have been implicitly extended to the whole star during the integration! The resulting difference will be accounted for by the function $\mathcal{R}$, such that:
\begin{equation}
    \label{eq:freqnum}
    \sigma^2 = \sigma^2_\mathrm{mod} + \mathcal{R} = \int_0^1\left(\sigma_c^2\,\xib^2+{\left(\frac{\partial\xib}{\partial t}\right)}^2\right)dt + \mathcal{R}.
\end{equation}

In this form, Eq.~\eqref{eq:freqnum} is particularly well-suited to the study of a frequency difference $\delta\sigma = \sigma^B - \sigma^A$ between two models A and B. Using the variational principle \citep{Chandrasekhar1964} in order to perturb Eq.~\eqref{eq:freqnum}, it merely provides at first order:
\begin{equation}
    \label{eq:intwave_p}
    \delta\sigma^2 = \delta\sigma_\mathrm{mod}^2+\delta\mathcal{R} \quad \mbox{with} \quad \delta\sigma_\mathrm{mod}^2 \equiv \int_0^1\delta_t(\sigma_c^2)\,\xib^2\,dt,
\end{equation} where $\delta_t$ designates a difference of profiles at identical $t$ (where $\delta$ stands for a simple difference of quantities) between the two models \citep[\PI]{JCD1997}. The advantage in using normalised variables becomes apparent: while no problem arises with the difference in profiles at a fixed $t$ value, it would have been different at a fixed $\tau$ value if ever $\tau_0^A\neq\tau_0^B$. 

Equation~\eqref{eq:intwave_p} satisfies the objective set at the end of Section~\ref{MOD}: it provides an explicit link between the perturbation of the oscillation frequencies $\delta\sigma^2$ and the perturbation of the structure through $\delta_t(\sigma_c^2)$. In this equation, $\delta \mathcal{R}$ characterises how closely the actual glitch $\delta\sigma^2$ is approximated by our modelled glitch expression $\displaystyle \delta\sigma_\mathrm{mod}^2$. We note that it is not necessary for $\sigma^2_\mathrm{mod}$ to be a good approximation of $\sigma^2$ for $\delta\mathcal{R}$ to be small. Indeed, it is only necessary for the assumptions to be verified in the regions where the structure differs between the models (since in the other regions $\delta_t(\sigma_c^2)$ would be strictly null in any case). 

\subsection{Comparison of glitch models with a numerically derived glitch \label{COMPARISON}}

Although Eq.~\eqref{eq:intwave_p} does meet the criteria we set ourselves, it does not constitute the only possible approximation of the glitch. In particular, we propose to compare Eq.~\eqref{eq:intwave_p} (to which we will refer with the index I) with the approximation of the glitch proposed by Eq.~\eqref{eq:dnu2_HG07} (index II):
\begin{align}
    \label{eq:dnuI}
    \delta\sigma_\mathrm{mod,~I}^2 ~&=~ \int_0^1\delta_t(\sigma_c^2)\,\xib^2\,dt, \\
    \label{eq:dnuII}
    \delta\sigma_\mathrm{mod,~II}^2 ~&=~2\sigma^2\frac{\delta\tau_0}{\tau_0}+\frac{\displaystyle\int_V \delta_x\Gamma_1 P ~{\left(\mathrm{div~}\vec{\upxi}\right)}^2\,dV}{\displaystyle \int_V \rho ~\vec{\upxi}^\mathsf{T}\vec{\upxi}\, dV}.
\end{align}

% GLITCH FIGURE --------------------------------------
   \begin{figure*}[!ht]
   \centering
   \includegraphics[width=\textwidth]{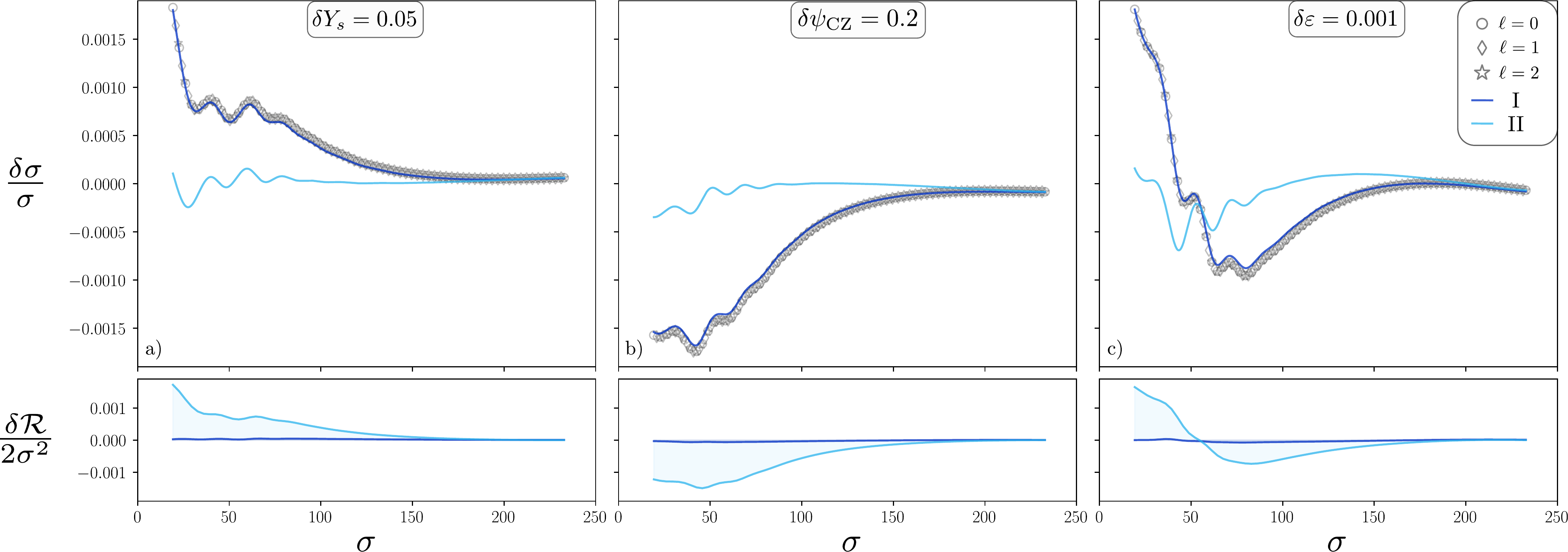}
   \caption{Comparison of numerically derived glitches $\delta\sigma/\sigma$ (black symbols, $0\leq\ell\leq 2$, $5\leq n\leq 70$) with both modelled expressions $\delta\sigma_\mathrm{mod,~I}^2/2\sigma^2$ (dark blue curves) and $\delta\sigma_\mathrm{mod,~II}^2/2\sigma^2$ (light blue curves) given by Eqs.~\eqref{eq:dnuI}-\eqref{eq:dnuII} for three differences in the parameters: (a) $\delta Y_s = 0.05$, (b) $\delta\psi_\mathrm{CZ} = 0.2$ and (c) $\delta\varepsilon = 0.001$. The bottom panels show the residual functions $\delta\mathcal{R}/2\sigma^2$ for both approximations.}
        \label{glitch_123.pdf}%
    \end{figure*}
% --------------------------------------------------------

In order to compare these two formulae, it is necessary to define the structural models A and B from which the glitch results. It may be noted that neither Eq.~\eqref{eq:dnuII} nor Eq.~\eqref{eq:dnuI} imposes a particular choice for the model A or B as long as it is possible to derive the required structural perturbations. Here we choose two instances of the structural models defined in \PI, $\mathcal{S}_{\vec{\uptheta}^\odot}(t)$ and $\mathcal{S}_{\vec{\uptheta}^\odot+\delta\vec{\uptheta}}(t)$, respectively parameterised by $\vec{\uptheta}^\odot = [Y_s^\odot = 0.255, \psi_\mathrm{CZ}^\odot = -5, \varepsilon^\odot = 0.015]^\mathsf{T}$ and $\vec{\uptheta}^\odot+\delta\vec{\uptheta}$. Such a situation can be seen as ideal, in the sense that the assumptions used to derive both glitch expressions I and II are verified by design (in particular the two models differ only in the ionisation region). Expressions I and II thus have all the reasons to be satisfactory approximations of $\delta\sigma^2$ (at first order) and this assertion is tested in Fig.~\ref{glitch_123.pdf}. The reference for our comparison, the ``true'' glitch, is merely the difference of the oscillation frequencies between the structures $\mathcal{S}_{\vec{\uptheta}^\odot}(t)$ and $\mathcal{S}_{\vec{\uptheta}^\odot+\delta\vec{\uptheta}}(t)$. The frequencies of each structure have been calculated numerically using \textit{InversionKit} \citep{Reese2016b} and appears in black symbols. This comparison is made for low degree modes $\ell \leq 2$ and a wide range of radial orders to show the asymptotic behaviour. Three glitches are considered (respectively in panels (a), (b) and (c)), corresponding to a difference in each parameter, the amplitude of which was chosen to cause a similar impact on the frequencies in terms of amplitude: $\delta\vec{\uptheta} = \left\{[0.05, 0, 0]^\mathsf{T}, [0, 0.2, 0]^\mathsf{T}, [0, 0, 0.001]^\mathsf{T}\right\}$. Finally, in order to better account for the departures, we have chosen to compare $\delta\sigma/\sigma$ with $\delta\sigma_\mathrm{mod}^2/2\sigma^2$ rather than $\delta\sigma^2$ with $\delta\sigma_\mathrm{mod}^2$ directly.

A first thing to note about the glitches calculated numerically is that they seem to share a unique relationship regardless of their degree, which is in line with our hypothesis stating that the displacement resulting from these low degree modes is almost radial in the perturbed region. Also, in addition to the clearly visible oscillation caused by the ionisation of helium, which gradually fades away at high frequencies, a slow-varying component is apparent. This component, which is visible in every panel, persists longer than that of helium and also oscillates much more slowly. A standard interpretation in terms of glitch tends to suggest an origin that is both more localised and superficial than that of helium. We believe that this is the signature of hydrogen ionisation, which has an impact on $\sigma_c^2$ as can be seen in Fig.~\ref{omegac2.pdf} at $t\sim0.02$ (this is to be compared with the impact of helium ionisation at $0.1<t<0.2$).

Regarding the modelled expressions, the glitch evaluated using Eq.~\eqref{eq:dnuI} (darker curves) provides a near perfect recovery of the ``true'' glitch in each situation. Considering the residual function $\delta \mathcal{R}/2\sigma_\mathrm{num}^2$, a subtle undershoot is visible which is to be expected from a first order approximation. In comparison, Eq.~\eqref{eq:dnuII} (lighter curves) provides poorer approximations, even though it can account for part of the oscillation (note that $\delta\sigma_\mathrm{mod,~II}^2/2\sigma_\mathrm{num}^2$ was corrected by an offset to better account for the asymptotic behaviour). However, by looking at the residuals, we see that by neglecting the pressure and inertia perturbations Eq.~\eqref{eq:dnuII} has not only discarded the slowly-varying component but also some of the fast oscillating part of the glitch.

\section{Estimating the properties of the ionisation region from the glitch \label{FIT}}

As stated above, it is worth noting that we have not assumed a specific shape for the structural perturbations up to Eqs.~\eqref{eq:dnuI}-\eqref{eq:dnuII}. For instance, we considered the structural perturbations $\delta_t\sigma_c^2$, and $\delta_x\Gamma_1/\Gamma_1$ as given to perform the comparisons in Fig.~\ref{glitch_123.pdf}. As suggested in Section~\ref{MOD} (and in Fig.~\ref{model_scheme.pdf}), our objective is now to combine the structural perturbation model $\delta\mathcal{S}_{\vec{\uptheta}, \delta\vec{\uptheta}}(t)$ with Eq.~\eqref{eq:dnuI} to obtain a parameterised model of the ionisation glitch. With such a model, we will investigate in this section how to infer the properties of the ionisation region $\delta\vec{\uptheta}$ from the glitch $\delta\sigma^2$.

\subsection{Constructing a glitch model from a structural perturbation model}

Used as is, Eq.~\eqref{eq:intwave_p} could allow us to find, knowing the shape of the glitch, the profile of $\delta_t\sigma_c^2$ by inversion. It seems likely in practice that this recovery would be patchy, however, especially given the number of frequencies available and the complexity of the cut-off frequency profile. As pointed out in Section~\ref{MOD}, we instead prefer to consider the shape of the perturbation as constrained by a parametric structural perturbation model $\delta\mathcal{S}_{\vec{\uptheta}, \delta\vec{\uptheta}}(t)$. This implies the following assumption: $\delta_t\sigma_c^2 \simeq \sigma_c^2(t,\vec{\uptheta}+\delta\vec{\uptheta})-\sigma_c^2(t,\vec{\uptheta})$. The resulting perturbation $\delta_t\sigma_c^2$ can thus be fully described using the pair $(\vec{\uptheta}, \delta \vec{\uptheta})$. In practice, the quantity $\vec{\uptheta}$ designates the value of the parameters in our reference model and is thus known by design. In contrast, $\delta\vec{\uptheta}$ accounts for the parameter differences that best explain the structural difference and must be inferred from the available data. For brevity, our glitch model will therefore be written as a function of $\delta\vec{\uptheta}$ only, although it depends on the chosen reference:
\begin{equation}
    \label{eq:glitch_mod}
    \delta\sigma^2 \simeq \delta\sigma^2_\mathrm{mod}(\delta\vec{\uptheta}) = \int_0^1\delta_t\sigma_c^2(\delta\vec{\uptheta})\,\xib^2\,dt.
\end{equation}
The quantity $\delta_t\sigma_c^2$ being perfectly defined as a function of $\delta\vec{\uptheta}$, this expression can be used as is to estimate the variation in parameters responsible for the glitch. In the following, however, we will mainly focus on the linearised form of this expression:
\begin{equation}
    \label{eq:glitch_mod_lin}
    \delta\sigma^2_\mathrm{mod}(\delta\vec{\uptheta}) \simeq \left(\vec{\nabla}_\uptheta \sigma^2_\mathrm{mod}\right)^\mathsf{T}\delta\vec{\uptheta}
\end{equation}with
\begin{equation}
    \label{eq:grad_mod}
    \vec{\nabla}_\uptheta \sigma^2_\mathrm{mod} \equiv \int_0^1\left(\vec{\nabla}_\uptheta\sigma_c^2\right)\,\xib^2\,dt.
\end{equation}

As discussed in Section~\ref{MOD}, the main trade-off between the use of this model compared to fully analytical forms of the glitch is its numerical cost. Whatever the value of $\delta\vec{\uptheta}$, it will be necessary to compute the model $\mathcal{S}_{\vec{\uptheta}+\delta\vec{\uptheta}}(t)$ in order to find the profile $\sigma_c^2(t,\vec{\uptheta}+\delta\vec{\uptheta})$ required to calculate Eq.~\eqref{eq:glitch_mod}. Equation~\eqref{eq:glitch_mod_lin} completely eliminates this constraint since it is only necessary to estimate $\vec{\nabla}_\uptheta \sigma^2_\mathrm{mod}$ once and for all, thus introducing in the process an error in $\mathcal{O}\left(\delta\vec{\uptheta}^\mathsf{T}\delta\vec{\uptheta}\right)$. We note that this linearisation assumes that the gradient $\vec{\nabla}_\uptheta \sigma^2_\mathrm{mod}$ only depends slightly on the reference model's parameters, thus making our expression \eqref{eq:glitch_mod} only dependent on $\delta\vec{\uptheta}$ and locally insensitive to $\vec{\uptheta}$ (thus justifying its omission in the glitch model). 

\subsection{The fitting method}

We now quantify the overall error on the recovery of $\delta\vec{\uptheta}$ from a glitch $\delta\sigma^2$ using Eq.~\eqref{eq:glitch_mod_lin}. 
For this purpose, we will consider the example of a glitch caused by the change of parameters $\delta\vec{\uptheta}_\mathrm{true} = [0.05,0.2,0.001]^\mathsf{T}$. Regarding Fig.~\ref{glitch_123.pdf}, this implies studying the glitch that results from combining the perturbation in each panel. The benefit from this example is twofold: firstly, it can reveal potential non-linearities as it is a non-negligible perturbation (each impact is of the order of the helium change $\delta Y_s = 0.05$, hence their combination constitutes a significant departure from the reference). Secondly, a closer inspection of Fig.~\ref{glitch_123.pdf} leads us to expect a certain degree of degeneracy between the effects of $\delta Y_s$ and $\delta \psi_\mathrm{CZ}$. Although the helium glitch is fairly similar in all panels (comparable amplitudes, periods and damping), these two parameters also cause a comparable hydrogen glitch but with opposite signs. Since we can thus expect a near-cancellation by combining these two contributions, it is interesting to see how much information can be recovered about the parameter differences from Eq.~\eqref{eq:glitch_mod_lin}.

From now on, $\delta\vec{\upsigma^2}$ will denote the vector of frequency shifts forming the glitch at all radial orders. The recovery will be done by means of a probabilistic approach, that is assuming that our parameter vector $\delta\vec{\uptheta}$ is a random variable and trying estimate its distribution given the observation of a certain glitch $\delta\vec{\upsigma^2}$. Such a random variable will be referred to as $\delta\vec{\uptheta}|\delta\vec{\upsigma^2}$ and its probability density function as the posterior distribution. Such an approach may seem like ``overkill'' for a linear problem like this one, but it is nevertheless essential to account for the potential degeneracies of this problem as well as the uncertainties on our estimate. 

The distribution $p(\delta\vec{\uptheta}|\delta\vec{\upsigma^2})$ is estimated with the help of the python library \textit{emcee}\footnote{\url{https://emcee.readthedocs.io}}, which implements MCMC methods for sampling distributions. Further details about this procedure and the underlying assumptions can be found in Appendix~\ref{AI1}. The synthetic data, $\delta\vec{\upsigma^2}$, generated for this purpose result from the oscillation frequency differences between the structures $\mathcal{S}_{\vec{\uptheta}^\odot}(t)$ and $\mathcal{S}_{\vec{\uptheta}^\odot+\delta\vec{\uptheta}}(t)$ as was done in the previous section. The parameter recovery can thus be seen as an ``inverse crime'', in the sense that the assumed ionisation region structure in Eq.~\eqref{eq:glitch_mod_lin} actually corresponds to the one used to calculate the synthetic glitch. It constitutes therefore an ideal framework that tests more the derivation of Eq.~\eqref{eq:intwave_p} and the subsequent linearisation \eqref{eq:glitch_mod_lin} than the relevance of the structural model itself. In contrast to Section~\ref{FREQ}, we considered here only purely radial modes with orders ranging from 11 to 27, in accordance with what modes we expect to observe. The standard deviation on the frequencies was set to ${s_{\vec{\mathrm{n}}}}=5.10^{-3}$, which would results in a $\sim 0.2~\mu$Hz dispersion for frequencies typical of a solar-like star. Although optimistic, this value is not unrealistic either: it is typically of the order of the mean uncertainty on the radial and dipolar modes in the \textit{Kepler} LEGACY sample \citep{Lund2017}.

\subsection{The resulting posterior distribution}

\begin{figure}
   \centering
   \includegraphics[width=9cm]{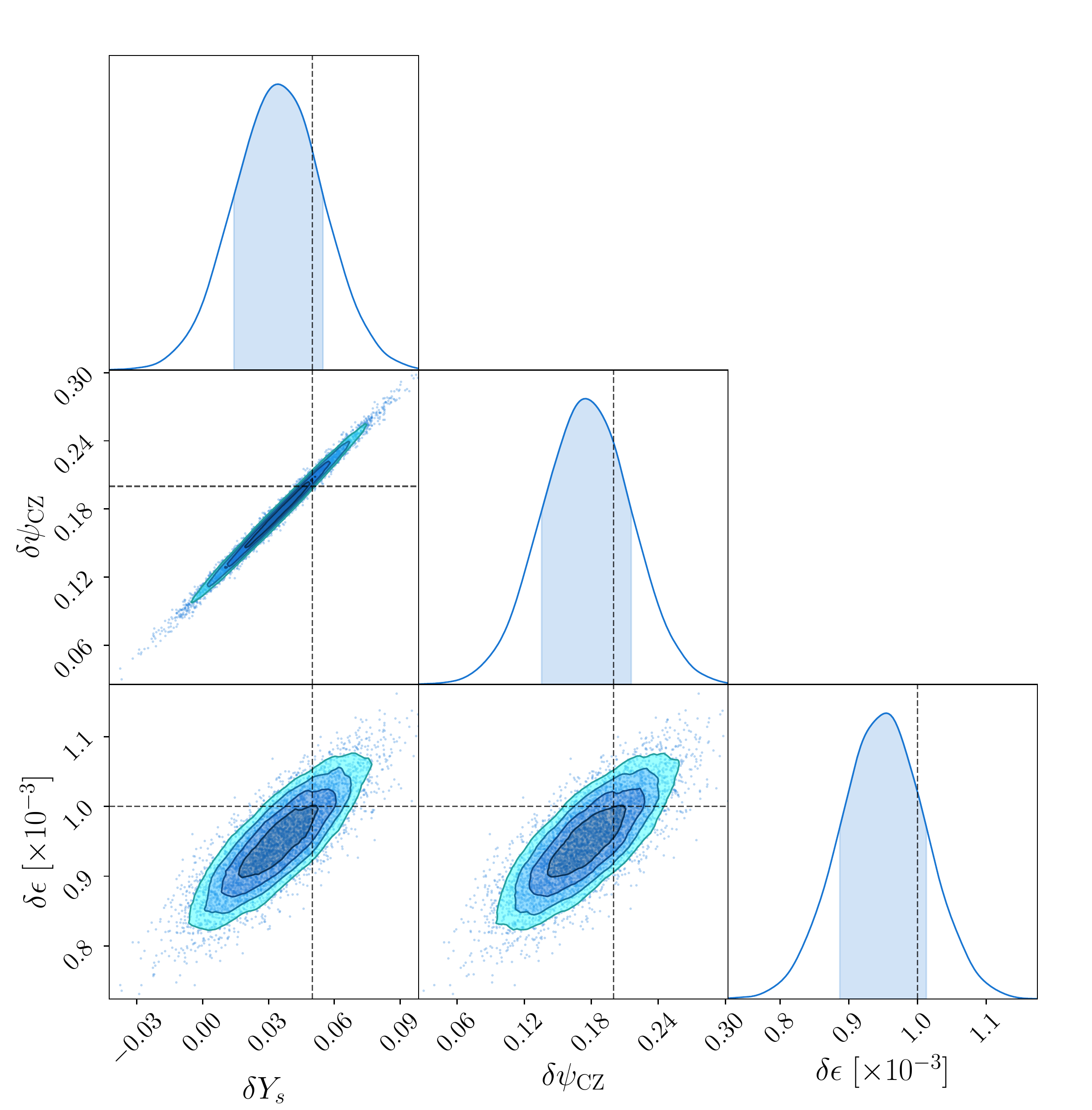}
   \caption{MCMC estimate of the distribution $p(\delta\vec{\uptheta}|\delta\vec{\upsigma^2})$. The dashed lines indicate the value of the true $\delta\vec{\uptheta}_\mathrm{true} = [0.05, 0.2, 1\times10^{-3}]^\mathsf{T}$ used to generate $\delta\vec{\upsigma^2}$ (the PDF gives at this point: $\ln p(\delta\vec{\uptheta}_\mathrm{true}|\delta\vec{\upsigma^2}) = 15.72$). The mode of this distribution is $\delta\vec{\uptheta}_\mathrm{MAP} = [0.036,0.178,0.951\times10^{-3}]^\mathsf{T}$. This plot was made with the python package \textit{ChainConsumer}\protect\footnotemark.}
        \label{mcmc_f.pdf}%
    \end{figure}

The resulting posterior distribution after applying the above procedure is shown in Fig.~\ref{mcmc_f.pdf} and summaries of this same distribution can be found in Table~\ref{tab:params}. The corner plot thus shows the univariate (two times marginalised) distributions on the diagonal as well as each bivariate (one time marginalised) distributions below, making the figure as a whole fairly representative of the underlying (3D) distribution.

Compared to the actual value of $\delta\vec{\uptheta}$ (shown with dashed lines), the most plausible value according to the PDF, $\delta\vec{\uptheta}_\mathrm{MAP}$, is somewhat underestimated ($\sim 30\%$ on $\delta Y_s$, $\sim 10\%$ on $\delta\psi_\mathrm{CZ}$ and $\sim 5\%$ on $\delta\varepsilon$) although still being fairly close. The PDF's extent is directly related to the noise level surrounding the oscillation frequencies ${s_{\vec{\mathrm{n}}}}$. Nevertheless, it is possible to draw some conclusions independently of a specific ${s_{\vec{\mathrm{n}}}}$ value: the estimate of the helium abundance seems to be the most volatile, followed by that of the electronic degeneracy, while the estimate of $\delta\varepsilon$ is the most stable. The reasons behind this statement are clearly apparent from Fig.~\ref{mcmc_f.pdf}. As expected, a strong degeneracy appears between $\delta Y_s$ and $\delta \psi_\mathrm{CZ}$ (the two parameters share a correlation coefficient of $0.98$) as they produce relatively similar effects on the glitch (cf. Fig.~\ref{glitch_123.pdf}). It should be noted that this degeneracy was already expected from the structure itself since both parameters impact the depth of the helium wells in the $\Gamma_1$ profile (\PI). This degeneracy especially increases the uncertainties on the helium abundance, making extreme values like $\delta Y_s = 0$ still relatively credible. It can also be noted that these two parameters also show a certain degree of correlation with $\delta\varepsilon$. This, unlike the previous degeneracy of physical origin, can be imputed to a lack of information in the data and in particular on the hydrogen component of the glitch. This very point is investigated in Appendix~\ref{AppendixII}, where a recovery of the parameters with additional low orders frequencies is presented.

Finally, it is tempting to judge the quality of the fit with a distance such as $\lVert\vec{\uptheta}_\mathrm{true}-\vec{\uptheta}_\mathrm{MAP}\rVert$ for instance, eventually defining a metric accounting for the distribution spread in each direction. However, it would be complicated for such a distance to account for strong correlations in the distribution, and in particular for the idea that ``points are probabilistically closer along a degeneracy than orthogonally''. A very natural way to proceed is to consider the value of the PDF at $\vec{\uptheta}_\mathrm{true}$, $\ln p(\delta\vec{\uptheta}_\mathrm{true}|\delta\vec{\upsigma^2})$, which penalises at the same time estimates that are too distant, too vague (because of the distribution's normalisation) while taking into account the degeneracies\footnote{We may note that this quantity is nothing more than the (negative) cross-entropy between the ``perfect'' Dirac distribution, $\delta\left(\delta\vec{\uptheta}-\delta\vec{\uptheta}_\mathrm{true}\right)$, and the posterior distribution $p(\delta\vec{\uptheta}|\delta\vec{\upsigma^2})$.}. In the present case, this metric gives the value $\ln p(\delta\vec{\uptheta}_\mathrm{true}|\delta\vec{\upsigma^2}) = 15.72$. This value on its own has only limited meaning (although it is clear that it should be as high as possible). It will nonetheless be interesting to compare it with other parameter recoveries in the next section.

% MCMC TABLE --------------------------------------
\begin{table}
    \begin{center}
    \caption{Comparison of the exact and fitted change in parameters.}
    \label{tab:params}
    \begin{tabular}{ccc}
        \hline
        \hline~\\[-3mm]
		\textbf{Parameters} & \textbf{``True'' values} & \textbf{Estimates}$^{~(*)}$ \\ 
		\hline~\\[-3mm]
		$\delta Y_s$ & $0.05$ & $0.034 \pm 0.020$ \\[0.1cm]
		$\delta\psi_\mathrm{CZ}$ & $0.2$ & $0.175^{+0.041}_{-0.039}$ \\[0.1cm]
		$\delta\epsilon$ & $1 \times 10^{-3}$ & $\left( 95.5^{+5.8}_{-6.8} \right) \times 10^{-5}$ \\ 
		\hline~\\[-1mm]
    \end{tabular}
    \end{center}
    \footnotesize
    $^{~(*)}$ : The values shown here are the distribution medians along with the $16^\mathrm{th}$ and $84^\mathrm{th}$ percentiles. We note that they differ slightly from the mode (cf Fig.~\ref{mcmc_f.pdf}) because of the distribution asymmetries.
    \normalsize
\end{table}
% --------------------------------------------------------

% MCMC FIGURE --------------------------------------
    \footnotetext{\url{https://samreay.github.io/ChainConsumer}}
% --------------------------------------------------------

% GLITCH SIM FIGURE --------------------------------------
   \begin{figure}[!ht]
   \centering
   \includegraphics[width=9cm]{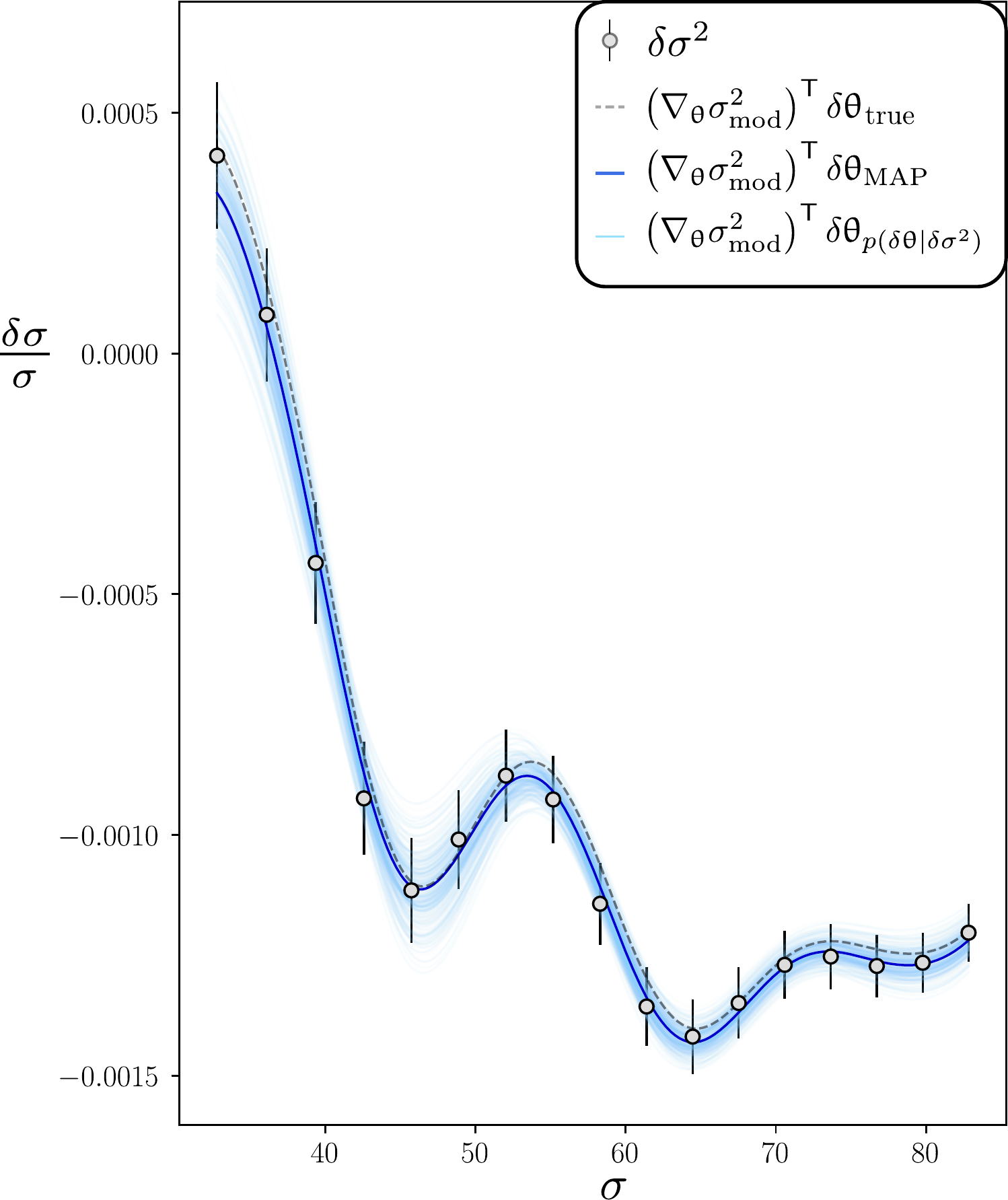}
   \caption{Comparison of the numerical glitch $\delta\sigma^2/2\sigma^2$ (circular symbols, $\ell = 0$ and $11\leq n \leq 27$) with the modelled glitch expression $\left(\vec{\nabla}_\uptheta \sigma^2_\mathrm{mod}\right)^\mathsf{T}\delta\vec{\uptheta}/2\sigma^2$ in the case of (1: grey dashed curve) $\delta\vec{\uptheta} = \delta\vec{\uptheta}_\mathrm{true}$ the actual value used to generate $\delta\vec{\upsigma^2}$, (2: dark blue curve) $\delta\vec{\uptheta} = \delta\vec{\uptheta}_\mathrm{MAP}$ the MAP estimate and (3: light blue curves) >300 different $\delta\vec{\uptheta}_{p(\delta\vec{\uptheta}|\delta\vec{\upsigma^2})}$ sampled from the posterior distribution $p(\delta\vec{\uptheta}|\delta\vec{\upsigma^2})$ (see Fig.~\ref{mcmc_f.pdf}).}
        \label{glitch_sampling_f.pdf}%
    \end{figure}
% --------------------------------------------------------

To visualise the posterior distribution in terms of frequency shifts, we show in Fig.~\ref{glitch_sampling_f.pdf} a comparison between the numerical glitch and our glitch model. We thus compare the model for the true value of the parameters, $\delta\vec{\uptheta}_\mathrm{true}$, the model for the most probable value of the parameters given the data, $\delta\vec{\uptheta}_\mathrm{MAP}$, and >300 curves drawn according to the posterior distribution on the parameters. These curves thus give an idea of how the spread of the distribution effectively propagates through the glitch models.

Regarding the appearance of the curves, it is not surprising to see that the synthetic glitch is fairly similar to the one shown in panel (c) of Fig.~\ref{glitch_123.pdf} (in the corresponding frequency range) since the helium and electronic degeneracy contributions to the glitch approximately cancel out. A first thing to note is that the model, even in this linear version, accurately reproduces the actual glitch for the true value of $\delta\vec{\uptheta}$ and thus justifies the linearisation in Eq.~\eqref{eq:glitch_mod_lin}. This curve is noticeably close to the one of the maximum a posteriori (MAP) estimate, $\delta\vec{\uptheta}_\mathrm{MAP}$, thus confirming the relevance of the retrieved solution. Also, the models drawn a posteriori account well for the uncertainties on the glitch based on the lighter curves. This make the helium degeneracy rather telling: glitch models that differ by more than $0.05$ in $\delta Y_s$ may sometimes barely be distinguished.

\section{Preliminary applications to seismic data \label{DISCUSSION}}

The results presented in the previous section suggest our method is able to recover most of the information on the ionisation region properties contained in the glitch, and this without needing any calibration on realistic models. Thus, despite the degeneracy affecting the reliability of our helium estimate, the procedure introduced seems to provide reasonable predictions for the parameters (as an indication, the MAP estimate on the total parameters would give $\vec{\uptheta}_\mathrm{MAP} = [ 0.291, -4.82, 1.60\times10^{-2}]^\mathsf{T}$ to be compared with $\vec{\uptheta}_\mathrm{true} = [ 0.305, -4.80,  1.60\times10^{-2}]^\mathsf{T}$). At this point, it is important to bear in mind that the recovery of parameters in Section~\ref{FIT} is carried out in an ideal framework, however. Although we have only considered an observable range of radial modes with an acceptable noise level, the difference between the two models under consideration is a perturbation of the ionisation region. By design, this implies the sole presence of the ionisation glitch in the frequency differences.

It is fairly easy to draw the limits of this case when confronted with actual seismic data\footnote{Note that it is not necessary to focus on potential changes to the Lamb frequency $\delta S_\ell^2$ as long as we only consider radial modes.}:
\begin{enumerate}
    \item By assuming $\gamma = \Gamma_1$, we neglected contributions of the Brunt-Väisälä frequency $\delta\mathcal{N}^2$ to the frequency shift $\delta\vec{\upsigma^2}$. Although they are indeed zero in the ionisation region, such differences could very well occur in a radiative region because of the $\mu$-gradient for instance \citep{Deheuvels2010,Noll2021}. Thus, the unexpected presence of a convective core in the observed star, or a deficient modelling of the chemical composition in this region can impact the differences in the oscillation frequencies. It should be noted, however, that the glitch model can easily be extended to the case $\delta\mathcal{N}^2\neq 0$ by simply replacing the cut-off frequency perturbation $\delta\sigma_c^2$ by the total acoustic potential perturbation $\delta\mathcal{V}$ \citep{Roxburgh1994b}.
    \item By restricting ourselves to the ionisation region we have neglected many of perturbations to the cut-off frequency occurring elsewhere. Others well-known contributions are the signature of the transition between the convective and radiative regions and surface effects. The first one is much deeper than the ionisation region for the stars we are interested in and thus manifests itself as a higher frequency (and lower amplitude) oscillation in the glitch \citep[e.g.][]{Monteiro1994,Houdek2007}. By far the most important in terms of amplitude, surface effects are at the origin of much slower components. They can be considered as the main source of ``contamination'' to the ionisation glitch fit and are one of the reasons for using second differences (cf. Eq.~\eqref{eq:d2nu}) instead of frequencies directly.
\end{enumerate}

These effects, if not treated properly, are likely to have a severe impact on our estimate of the properties of the ionisation region. In the remainder of this section, we will explore two approaches for applying our method despite the presence of interfering components in the data. The first is very common and seeks to reduce these components by combining frequencies and the second attempts to fit the frequencies directly using Gaussian processes.

\subsection{Using frequency combinations as a diagnostic \label{diagnostic}}

Considering a diagnostic, $\vec{\mathrm{y}}$, expressed as a linear combination of the squared frequencies, $\vec{\upsigma^2}$, we write:
\begin{equation}
    \label{eq:def_diff}
    \vec{\mathrm{y}} = \mathrm{C}\vec{\upsigma^2}
\end{equation} with $\mathrm{C}$ a matrix for which we can assume a general form at this point. We note here that the diagnostic introduced above differs slightly from those usually used, which are written as a combination of the frequencies directly. Using squared frequencies brings a more intuitive understanding in the formalism introduced so far, while giving identical results as confirmed by supplementary numerical tests. The general idea, and a fortiori the elimination of the most superficial components when $\mathrm{C}$ is the discrete derivative operator, remains very similar.

Then, from Eqs.~\eqref{eq:glitch_mod_lin} and~\eqref{eq:grad_mod}, it follows immediately from Eq.~\eqref{eq:def_diff} that
\begin{equation}
    \label{eq:diff_glitch_mod_lin}
    \delta \vec{\mathrm{y}} \simeq \left(\vec{\nabla}_\uptheta \vec{\mathrm{y}}_\textrm{mod}\right)^\mathsf{T}\delta\vec{\uptheta}
\end{equation}with
\begin{equation}
    \vec{\nabla}_\uptheta \vec{\mathrm{y}}_\textrm{mod} \equiv \int_0^1\left(\vec{\nabla}_\uptheta\sigma_c^2\right)\,\left(\mathrm{C}\Vec{\upeta^2}\right)^\mathsf{T}\,dt,
\end{equation}$\Vec{\upeta^2}$ being the vector composed of all the normalised eigenfunctions $\xib^2$. Note that in this vector form (accounting for all frequencies at once) $\vec{\nabla}_\uptheta \vec{\mathrm{y}}_\textrm{mod}$ designates a matrix, which is very similar to 
\begin{equation}
    \vec{\nabla}_\uptheta \vec{\upsigma^2}_\textrm{mod} \equiv \int_0^1\left(\vec{\nabla}_\uptheta\sigma_c^2\right)\,\left(\Vec{\upeta^2}\right)^\mathsf{T}\,dt,
\end{equation} the vector form of Eq.~\eqref{eq:grad_mod}. In fact, it refers to the decomposition of the same function, $\vec{\nabla}_\uptheta\sigma_c^2$, but on different kernels, namely $\mathrm{C}\Vec{\upeta^2}$ instead of $\Vec{\upeta^2}$. While $\Vec{\upeta^2}$ designates oscillating functions of constant amplitude in the major part of the star, the combinations $\mathrm{C}\Vec{\upeta^2}$ can take more general envelopes. In the specific case where $\mathrm{C} = \mathrm{D}^k$, with $\mathrm{D}^k$ the discrete derivative operator to a given power $k$, this envelope varies as $\sin^k(\pi t)$ independently of the frequency $\sigma$ \citep{Ballot2004}. This combination can thus be seen as enhancing the signal of $\vec{\nabla}_\uptheta\sigma_c^2$ at $t = t_1$ compared to $t = t_2$ by a factor of $(\sin(\pi t_1)/\sin(\pi t_2))^k$. With regard to the concerns raised above, the main advantage of this method clearly appears when it comes to getting rid of the surface effects. Taking the example of a ``contamination'' at $t_\mathrm{surf}\sim0.01$ and a helium glitch at $t_\mathrm{He}\sim0.15$, the signal of the latter is comparatively amplified by a factor $>200$ for $k=2$ (corresponding to a second derivative)! The same reasoning, however, also holds for the hydrogen component, for which $t_\mathrm{H}\sim 0.02$ thereby getting completely removed. It is therefore important to assess the impact of this loss of information in our fit, especially since we expect this component to be essential when recovering the different parameters.

% MCMC FIGURE --------------------------------------
   \begin{figure}
   \centering
   \includegraphics[width=9cm]{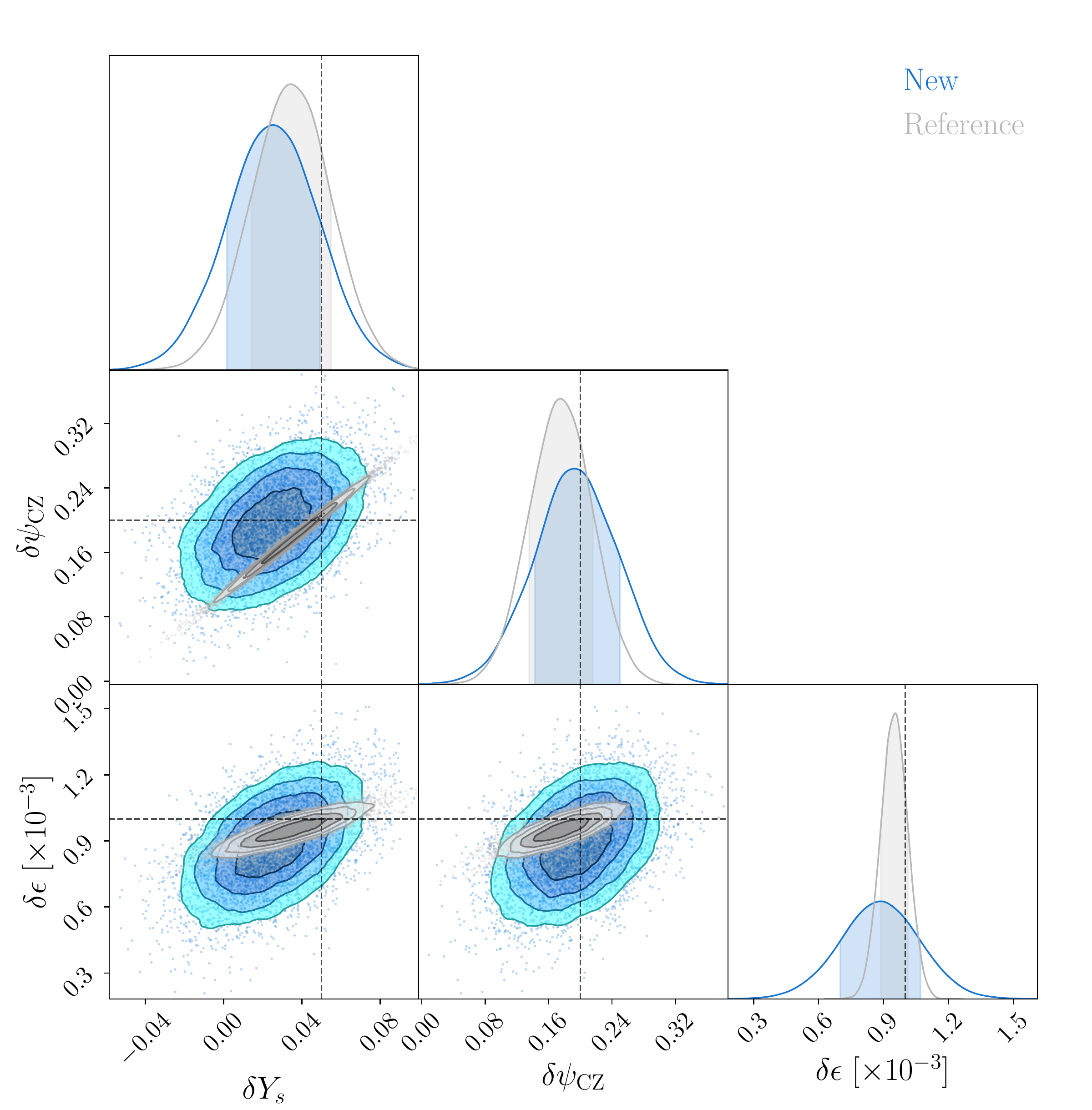}
   \caption{MCMC estimate of the distribution $p(\delta\vec{\uptheta}|\delta \vec{\mathrm{y}})$ in the case $\vec{\mathrm{y}} = \mathrm{D}^2\vec{\upsigma^2}$ (in blue) compared to the case $\vec{\mathrm{y}} = \vec{\upsigma^2}$ shown in Fig.~\ref{mcmc_f.pdf}. The dashed lines indicate the true value of $\delta\vec{\uptheta}_\mathrm{true} = [0.05, 0.2, 1\times10^{-3}]^\mathsf{T}$ used to generate $\delta \vec{\mathrm{y}}$ (the PDF gives at this point: $\ln p(\delta\vec{\uptheta}_\mathrm{true}|\delta \vec{\mathrm{y}}) = 12.17$). The mode of this distribution is $\delta\vec{\uptheta}_\textrm{MAP} = [0.026,0.195,0.888\times10^{-3}]^\mathsf{T}$.}
        \label{mcmc_d2f.pdf}%
    \end{figure}
% --------------------------------------------------------

% GLITCH SIM FIGURE --------------------------------------
   \begin{figure}[!ht]
   \centering
   \includegraphics[width=9cm]{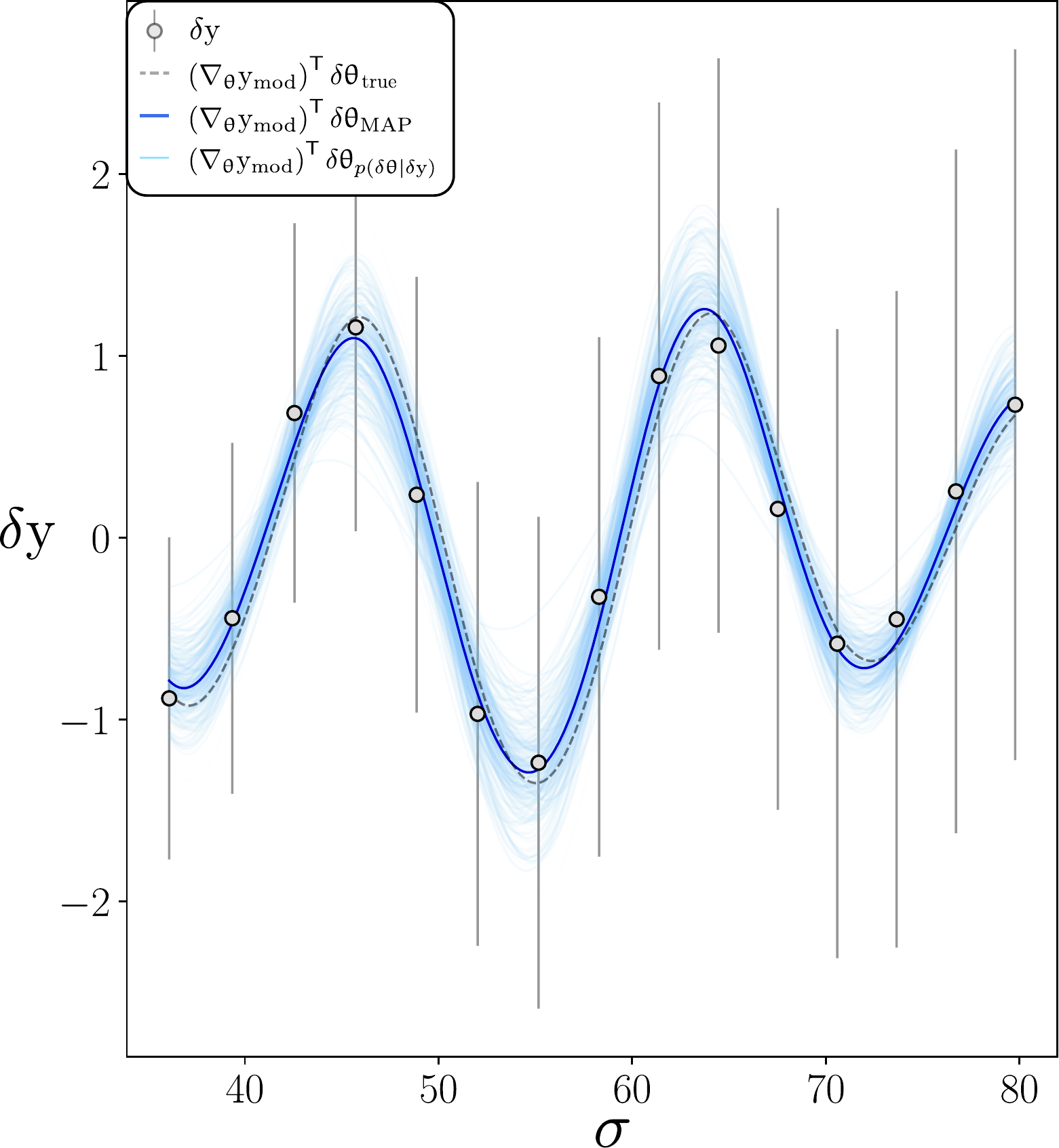}
   \caption{Same as Fig.~\ref{glitch_sampling_f.pdf} but for the the synthetic diagnostic $\delta \vec{\mathrm{y}}$ and the modelled expression $\left(\vec{\nabla}_\uptheta \vec{\mathrm{y}}_\textrm{mod}\right)^\mathsf{T}\delta\vec{\uptheta}$.}
        \label{glitch_sampling_d2f.pdf}%
    \end{figure}
% --------------------------------------------------------

Thus, we performed a similar analysis to that conducted in Section~\ref{FIT}, this time aiming to fit the perturbation of the second differences $\delta\vec{\mathrm{y}} = \delta(\mathrm{D}^2\vec{\upsigma^2})$ instead of that of the frequencies $\delta\vec{\upsigma^2}$ directly. The method is completely analogous to that described above, and some details can be found in Appendix~\ref{AI2}. For the exact same value of $\delta\vec{\uptheta}_\mathrm{true} = [0.05, 0.2, 1\times10^{-3}]^\mathsf{T}$, Fig.~\ref{mcmc_d2f.pdf} gives a MCMC sampling of the posterior distribution $p(\delta\vec{\uptheta}|\delta \vec{\mathrm{y}})$ using the diagnostic $\delta \vec{\mathrm{y}}$. Compared to the reference distribution $p(\delta\vec{\uptheta}|\delta\vec{\upsigma^2})$, found in Fig.~\ref{mcmc_f.pdf} and shown in greyscale here, this new PDF is notably more spread out. This is not a surprise, since the elimination of the slowly varying component can only result in a loss of information; nevertheless it is worth quantifying it. We see that an important loss of information concerns the parameter $\delta\varepsilon$, which can be seen in the spreading of its marginal distribution. As mentioned in Section~\ref{FIT}, with the compensation of the contributions of $\delta Y_s$ and $\delta\psi_\mathrm{CZ}$, it is expected that the hydrogen component mainly constrains $\delta\varepsilon$. Thus, its absence here appears most clearly on this last parameter. 

Besides this, the shape of the distribution significantly changed on the $\delta Y_s$ versus $\delta \psi_\mathrm{CZ}$ panel, with the clear degeneracy between the two parameters being now more diffuse. Actually, without the reference distribution to compare with, we might have concluded that the degeneracy was removed. However, the comparison makes it apparent that it is rather the presence of the helium oscillation alone (i.e. without the contribution of hydrogen) that allows more pairs $(\delta Y_s, \delta \psi_\mathrm{CZ})$ to be credible. Thus, the physically-based degeneracy between $\delta Y_s$ and $\delta \psi_\mathrm{CZ}$ seems in a sense ``resolved'' when sufficient information is available, while getting hidden in the uncertainty when knowledge on the slowly varying component starts missing. 

Finally, the mode of distribution, $\delta\vec{\uptheta}_\textrm{MAP} = [0.026,0.195,0.888\times10^{-3}]^\mathsf{T}$, has changed compared to the previous case, although not drastically regarding the extent of the distribution (we may note however that the most plausible value of $\delta Y_s$ is underestimated by $\sim 50\%$ compared to the true value). As mentioned earlier, however, a more appropriate way to judge the quality of the fit is through the value of the distribution in $\delta\vec{\uptheta}_\textrm{true}$, which gives $\ln p(\delta\vec{\uptheta}_\mathrm{true}|\delta \vec{\mathrm{y}}) = 12.17$. This metric notably tells us that $p(\delta\vec{\uptheta}_\mathrm{true}|\delta \vec{\mathrm{y}})$ is $\sim 35$ times lower than $p(\delta\vec{\uptheta}_\mathrm{true}|\delta \vec{\upsigma^2})$ and thus quantifies the deterioration of the fit by using the second differences as a diagnostic. 

We once more compared our model provided by Eq.~\eqref{eq:diff_glitch_mod_lin} and the diagnostic obtained from numerical frequency deviations in Fig.~\ref{glitch_sampling_d2f.pdf}. This time a single oscillatory component, that of helium, appears as expected. A first observation stands out, namely the considerable amplification of the uncertainty due to the linear combination. The errors shown in the figure are given by the square root  of the diagonal coefficients of $\Sigma_{\mathrm{y}}$, the covariance matrix, for lack of a better representation. Thus, while being visually dominant, these error bars are not necessarily representative of the uncertainties on the diagnostic $\vec{\mathrm{y}}$ as they do not account for the correlated part of the noise (which helps to reduce some of the overall uncertainty). We may also note the proximity of the curves given by $\left(\vec{\nabla}_\uptheta \vec{\mathrm{y}}_\textrm{mod}\right)^\mathsf{T}\delta\vec{\uptheta}$ and $\left(\vec{\nabla}_\uptheta \vec{\mathrm{y}}_\textrm{mod}\right)^\mathsf{T}\delta\vec{\uptheta}_\mathrm{MAP}$ despite their difference in terms of the parameters. More generally, many curves drawn from $p(\delta\vec{\uptheta}|\delta \vec{\mathrm{y}})$ are of comparable amplitude, whereas some of them differ by more than $0.1$ in $\delta Y_s$ for instance.

Overall, we can see that the use of a diagnostic, while it mitigates the contamination to the ionisation glitch such as the one caused by surface effects, might hinder our ability to recover the solution. Even in the ideal case presented here (in the absence of a contamination), finding a way to preserves the hydrogen component appears essential. Thus, we will see in the following section an approach to fit directly the ionisation glitch even in the presence of additional components.

\subsection{Using frequencies directly}

In this paragraph, we will take the explicit example of a contamination by surface effects and therefore consider a direct approximation of $\delta\vec{\upsigma^2}$ by relying on the following model:
\begin{equation}
    \label{eq:df2_ion_surf}
    \delta \vec{\upsigma^2}_\mathrm{mod}(\delta\vec{\uptheta}, \vgp) = \delta \vec{\upsigma^2}_\mathrm{ion}(\delta\vec{\uptheta}) + \delta \vec{\upsigma^2}_\mathrm{surf}(\vgp),
\end{equation} where $\delta \vec{\upsigma^2}_\mathrm{ion}(\delta\vec{\uptheta})$ is given by Eq.~\eqref{eq:glitch_mod_lin} and $\delta \vec{\upsigma^2}_\mathrm{surf}(\vgp)$ designates a parameterised model of the surface effects. Expressing such a model explicitly is by no means straightforward, especially since what we call "surface effects" depends more on what is included (or not) in our reference structural model rather than on a well-defined physical phenomenon. It is thus clear that the model describing the corresponding frequency shifts must be somewhat flexible. Meanwhile, this model must satisfy constraints, which are:
\begin{enumerate}
    \item To vary on sufficiently large scales. By definition, these effects include the most superficial deviations from the structure.
    \item To be much weaker at low than at high frequencies. Indeed, the acoustic modes are becoming more and more sensitive to the surface effects with increasing radial orders.
\end{enumerate}

\begin{figure}
   \centering
   \includegraphics[width=9cm]{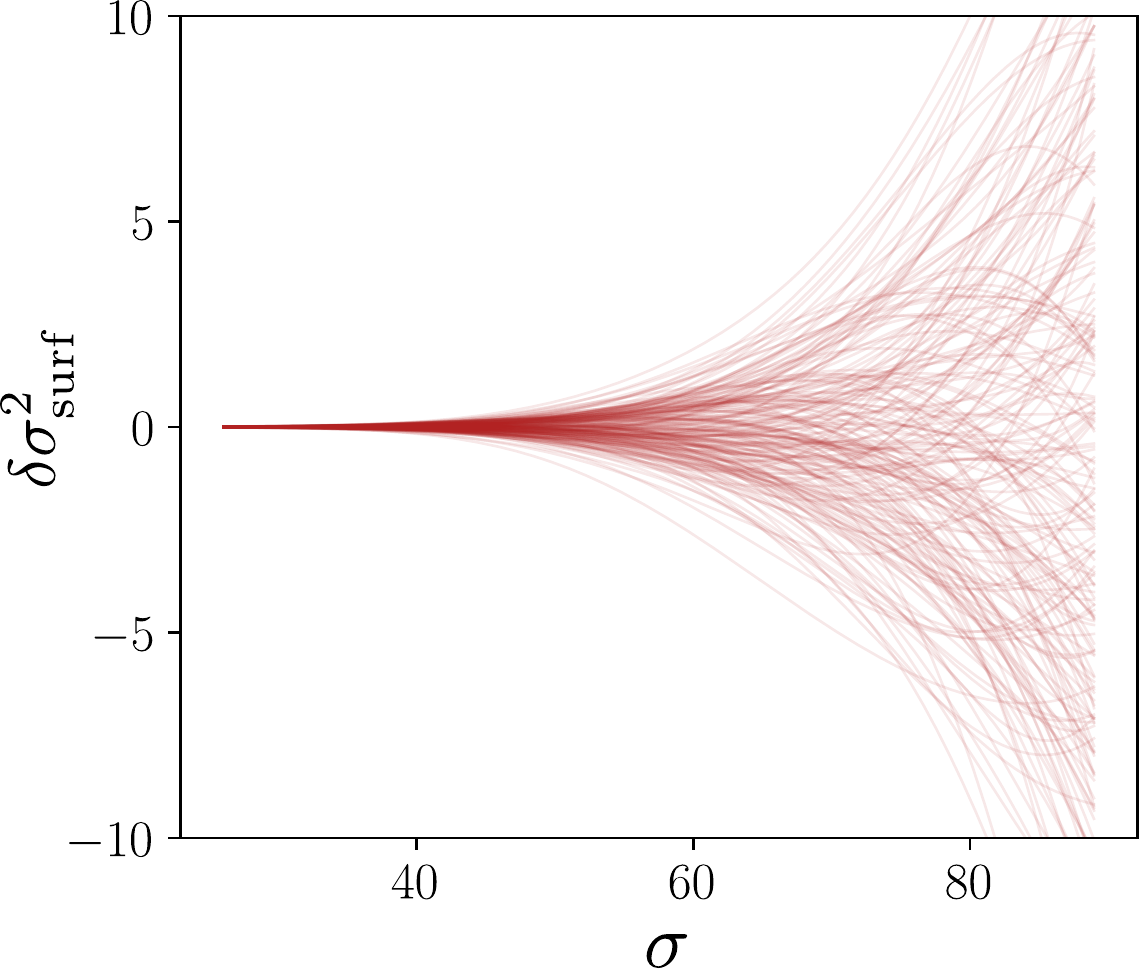}
   \caption{Example of $200$ realisations of the Gaussian process $\delta\sigma^2_{\mathrm{surf},\vgp}(\sigma)$ (cf. Eqs.~\eqref{eq:def_gp}-\eqref{eq:ksurf}) for a fixed value of $\vgp = [\log \beta^2, \log M]^\mathsf{T} = [50, 6]^\mathsf{T}$.}
        \label{gp_sampling.pdf}%
    \end{figure}

To satisfy these conditions, we have chosen here to model these surface effects by a Gaussian process (GP) whose parameterisation by $\vgp$ is specified in Appendix~\ref{AI3}. To illustrate the possibilities offered by this GP, Fig.~\ref{gp_sampling.pdf} shows $200$ realisations of the process given by Eqs.~\eqref{eq:def_gp}-\eqref{eq:ksurf} for a fixed set of values for $\vgp$. As desired, the generated functions are indeed of low amplitude at low frequencies and exhibit a polynomial envelope. They also show a certain flexibility and generally end up deviating from a simple power law at higher frequencies, which allows them to mimic a more comprehensive behaviour. It is also clear that many of the functions in this sample do not correspond to a plausible observation of surface effects. Without observational constraints, this sample only provides us with an \textit{a priori} view of the functions that the process generates.

% GP SAMPLING FIGURE -------------------------------------
% --------------------------------------------------------

\begin{figure}
   \centering
   \includegraphics[width=9cm]{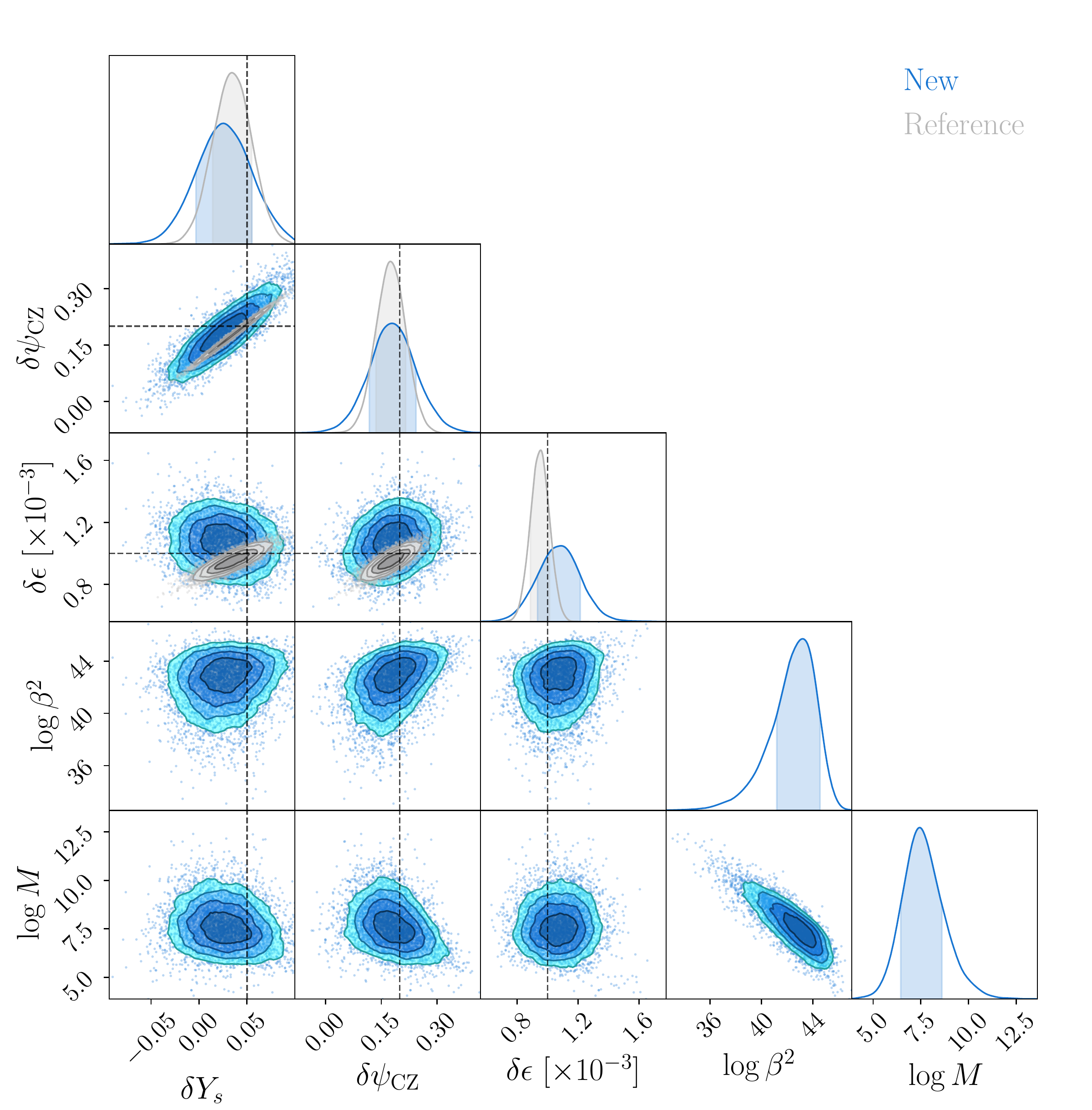}
   \caption{MCMC estimate of the distribution $p(\delta\vec{\uptheta},\vgp|\delta \vec{\upsigma^2})$ (in blue) compared to the reference without any contamination found in Fig.~\ref{mcmc_f.pdf} (in grey). The dashed lines indicate the value of $\delta\vec{\uptheta}_\mathrm{true} = [0.05, 0.2, 1\times10^{-3}]^\mathsf{T}$ used to generate $\delta \vec{\upsigma^2}$ (the marginalised PDF over $\vgp$ gives at this point: $\ln p(\delta\vec{\uptheta}_\mathrm{true}|\delta\vec{\upsigma^2}) = 12.67$). The mode of this distribution is given by $\delta\vec{\uptheta}_\textrm{MAP} = [0.027,0.178,1.051\times10^{-3}]^\mathsf{T}$ and $\vgp_\textrm{MAP} = [42.8, 7.63]^\mathsf{T}$}
        \label{mcmc_gp.pdf}%
    \end{figure}

On a practical level, the use of GPs is implemented via the python library \textit{george}\footnote{\url{https://george.readthedocs.io}} while the distribution sampling is carried out as in Section \ref{FIT}. Naturally, we need to synthetically generate surface effects this time in order to test the method in a controlled framework. We followed the prescription of \cite{Sonoi2015}, to generate ad-hoc surface effects that we would expect to find from a solar-like star. Based on their scaling relations and taking into account our non-dimensionalisation, we obtained the following expression:
\begin{equation}
    \delta\sigma^2_\mathrm{surf,~true}(\sigma) = 2\sigma\delta\sigma_\mathrm{surf,~true}(\sigma) = -0.6\times\sigma\left(1-\frac{1}{1+(\sigma/60)^5}\right).
\end{equation}
This artificial frequency shift was added to the glitch studied in Section \ref{FIT} in order to generate the ``observed'' frequency deviation $\delta \vec{\upsigma^2}$. The probability distribution of the parameters given this data realisation, $p(\delta\vec{\uptheta},\vgp|\delta \vec{\upsigma^2})$, is shown in Fig.~\ref{mcmc_gp.pdf} (in blue scale) and compared with the one obtained in Fig.~\ref{mcmc_f.pdf} in the absence of contamination (in grey scale). As was the case in the previous section, the distribution has expanded, due on the one side to the uncertainty caused by surface effects and on the other side to the fixed amount of information that is now shared by 5 parameters. In fact the marginalised distribution over $\vgp$, visible in the 6 upper panels, shows many similarities with $p(\delta\vec{\uptheta}|\delta \vec{\mathrm{y}})$ in Fig.~\ref{mcmc_d2f.pdf}. In particular, we observe the more diffuse degeneracy between $\delta Y_s$ and $\delta\psi_\mathrm{CZ}$ as well as the expansion of the distribution on the previously best constrained parameter, $\delta\varepsilon$. These effects occur to a lesser degree, however, as indicated by the value of $\ln p(\delta\vec{\uptheta}_\mathrm{true}|\delta\vec{\upsigma^2}) = 12.67$ being larger than $\ln p(\delta\vec{\uptheta}_\mathrm{true}|\delta\vec{\mathrm{y}})$. In some ways, the fact that this method provides a better parameter recovery in the presence of synthetic surface effects than using second differences without any contamination emphasises the benefits of the procedure. Also, as was done with the contamination-free fit, we investigate the improvement of this recovery with the use of additional frequencies in Appendix~\ref{AppendixII}.

\begin{figure}[!ht]
   \centering
   \includegraphics[width=9cm]{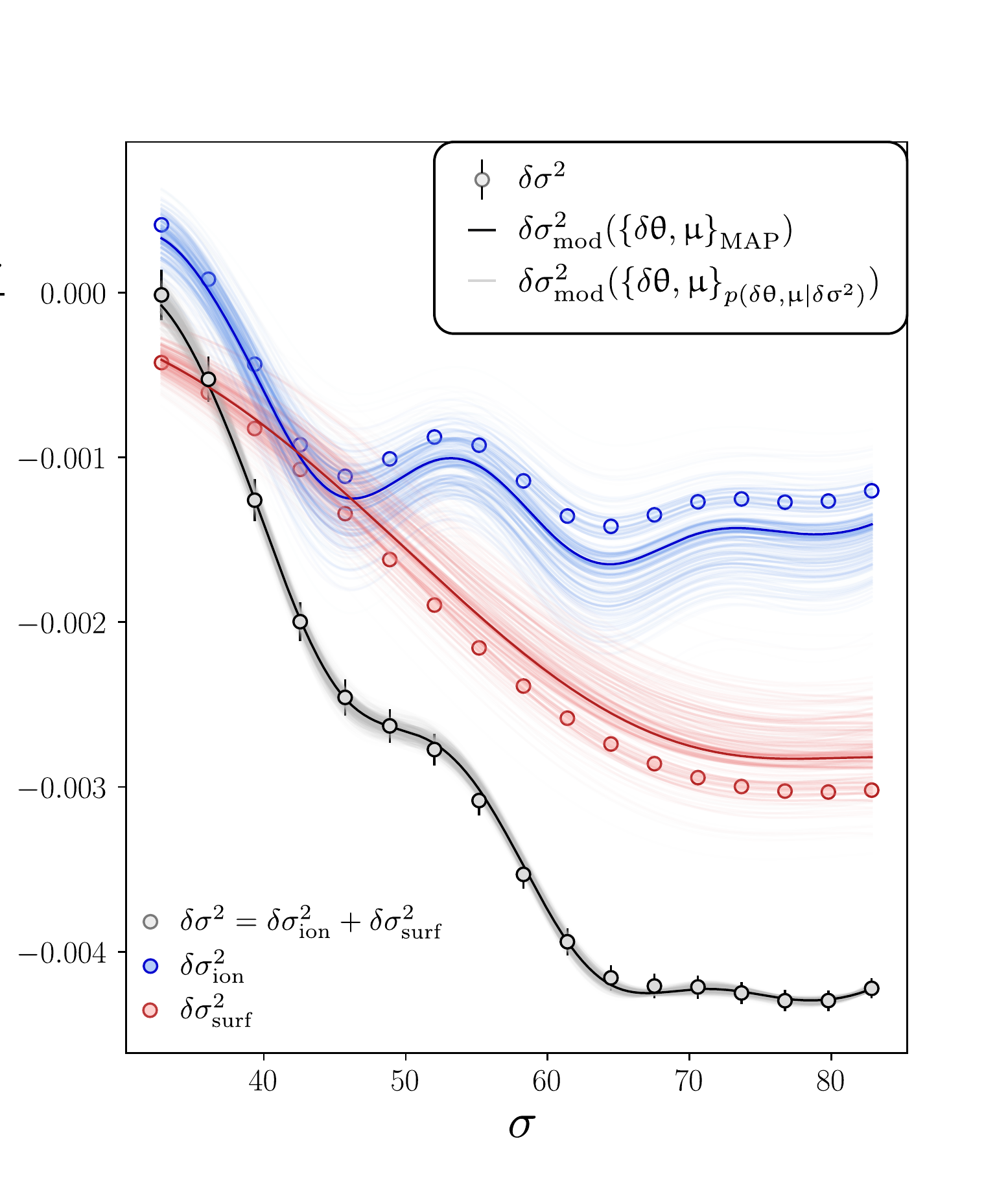}
   \caption{Comparison of the synthetic frequency shift $\delta \vec{\upsigma^2}$ (black symbols) resulting from the ionisation glitch $\delta \vec{\upsigma^2}_\mathrm{ion}$ (blue symbols), and a contamination $\delta \vec{\upsigma^2}_\mathrm{surf}$ (red symbols) with the modelled glitch expression $\delta \vec{\upsigma^2}_\mathrm{mod}(\delta\vec{\uptheta}, \vgp) = \delta \vec{\upsigma^2}_\mathrm{ion}(\delta\vec{\uptheta}) + \delta \vec{\upsigma^2}_\mathrm{surf}(\vgp)$ (respectively shown with black, blue and red curves). For each component the darker curve corresponds to the MAP estimate while the lighter ones result from the sampling of $p(\delta\vec{\uptheta},\vgp|\delta \vec{\upsigma^2})$.}
        \label{glitch_sampling_gp.pdf}%
    \end{figure}

Once again, a more visual representation of this distribution is available in Fig.~\ref{glitch_sampling_gp.pdf} through realisations of the models $\vec{\upsigma^2}_\mathrm{mod}(\delta\vec{\uptheta}, \vgp) = \delta \vec{\upsigma^2}_\mathrm{ion}(\delta\vec{\uptheta}) + \delta \vec{\upsigma^2}_\mathrm{surf}(\vgp)$, $\delta \vec{\upsigma^2}_\mathrm{ion}(\delta\vec{\uptheta})$ and $\delta \vec{\upsigma^2}_\mathrm{surf}(\vgp)$ according to the distribution $p(\delta\vec{\uptheta},\vgp|\delta \vec{\upsigma^2})$ (light curves). Figure~\ref{glitch_sampling_gp.pdf} also allows us to assess the magnitude of the surface effects (red symbols) compared to the glitch. In particular, their amplitude is three times higher at high frequencies, largely impacting the resulting frequency shift (black symbols) when compared to the ionisation glitch (blue symbols). Despite this, it is not surprising that the model in its entirety (black curves) manages to adequately reproduce this frequency shift considering the added flexibility it is given. Having said this, it is far more interesting to see what the model assumes about $\delta \vec{\upsigma^2}_\mathrm{surf}(\vgp)$ and especially $\delta \vec{\upsigma^2}_\mathrm{ion}(\delta\vec{\uptheta})$ (red and blue curves respectively) based on the sole observation of $\delta\vec{\upsigma^2}$. Concerning the surface effects model, we can see in particular that the mean of the GP has clearly changed in comparison with the a priori process (cf Fig.~\ref{gp_sampling.pdf}). All these curves do look like plausible realisations of surface effects (at least in the light of those we artificially imposed) and overall only slightly underestimate their amplitude. The shape of the glitch is also fairly well guessed, although the most likely model gradually deviates from it at high frequencies. We note that we obtain constraints not only on the fast oscillation but also on the slower trend of the glitch. Thus, some of the information on the hydrogen component is recovered and more can be retrieved if it is possible to obtain even lower radial orders (cf. Appendix~\ref{AppendixII}). The envelope formed by the sampling is naturally larger than in Fig.~\ref{glitch_sampling_f.pdf} and thus reflects the uncertainty introduced by the added contamination.

% MCMC GP FIGURE -----------------------------------------
% --------------------------------------------------------

% GLITCH SAMPLING GP FIGURE -----------------------------
% --------------------------------------------------------

\section{Conclusion \label{CONCLUSION}}

In the present paper, we develop a method for estimating the properties of the ionisation region from the glitch caused by rapid structural variations in this very region. In particular, the method presented here seeks to overcome the dependence on realistic stellar models caused by calibration in order to provide a truly independent estimate of these properties. These include both the helium abundance and other physical quantities that can have a significant impact on the oscillation frequencies such as the electronic degeneracy parameter or the extent of the ionisation region.

This paper takes as a starting point \PI~and combines the structural perturbations of the ionisation zone presented there with the perturbation of a wave equation. We considered for that a framework suited to the ionisation region (purely radial oscillations in an isentropic region) and verified that the resulting formula is able to correctly reproduce glitches produced by various structural perturbations. Although it only involves the perturbation of the (normalised) cut-off frequency $\delta\sigma_c^2$, this formulation is more accurate than common glitch expressions that only consider the perturbation of the first adiabatic exponent $\delta\Gamma_1$. This gain in accuracy is especially useful for us: the resulting glitch model is thus able to exploit the information contained in the fast oscillation caused by the helium ionisation but also in the slow trend accompanying that of hydrogen. By using the work conducted in \PI, this information can directly be expressed in terms of parameters $\delta\vec{\uptheta} = [\delta Y_s,\delta \psi_\textrm{CZ},\delta \varepsilon]^\mathsf{T}$, respectively representing a difference in helium abundance, electronic degeneracy and extent of the ionisation region (from a fixed reference).

This parameter recovery has been quantified through Bayesian inference in the context of limited (observable radial modes) and noisy synthetic data. By design, the ``observed'' glitch is generated based only on structural differences localised in the ionisation region, and therefore constitutes a robust reference for evaluating the procedure's performances in an ideal case (without any external contamination). It was thus verified that the properties at the origin of the glitch can mostly be recovered. Specifically, the best constrained parameter is the region's extent, followed by electronic degeneracy, while the most variable estimate concerns the helium abundance. At the origin of this uncertainty is a degeneracy between $\delta Y_s$ and $\delta \psi_\textrm{CZ}$ which was already expected in \PI~and which particularly extends over the helium abundance. Unlike the correlations between the other parameters which disappear when more data are available (see Appendix~\ref{AppendixII}), this degeneracy has a physical origin and does not seem to be specifically inherent to this approach. Indeed, both parameters have (locally) similar impacts within the star's structure itself, which fundamentally makes it difficult to disentangle them. Thus, an increase in the number of observed frequencies or even a decrease in the considered noise level does not make this degeneracy disappear but limits its extent.

We then considered how to extend the procedure to the non-ideal (but more realistic) case of a contamination of the observed glitch by an additional component, typically surface effects. A first possibility is to extend the formalism to other diagnostics such as frequency combinations. While this adaptation does not raise any fundamental problems, a strong degradation of the parameter recovery is already observed in the ideal case. We attribute this deterioration to the loss of information on the slow trend of the glitch. Indeed, while frequency combinations allow the amplitude of a slowly varying contamination to be mitigated, they also remove all trace of the glitch component resulting from the ionisation of hydrogen. We therefore considered a second possibility in order to preserve this component despite the contamination by exploiting Gaussian processes. The latter allows us to introduce an additional degree of flexibility in the model and takes into account additional components not explicitly defined while introducing a limited number of additional parameters (2 in our case). A deterioration appears compared to the ideal case, but mainly due to the impact of the surface effects (the amplitude of which is a few times higher than that of the glitch at high frequencies) and the fact that a fixed amount of information in the data is now used to constrain additional parameters. It was thus possible to graphically verify that the slowly-varying component was partially recovered and the estimate of surface effects was plausible. We note that this fit provides better results (in the sense of cross-entropy with the target distribution) than a second differences fit in the absence of contamination. 

This work now needs to be extended to the analysis of actual data and account for the accuracy and uncertainty on a retrieval of the ionisation region properties. To this end, we point out the benefit of combining a glitch model with a Gaussian process. The number of options offered by GPs is immense and the results presented here show only one particular attempt based on weak assumptions. It is clear that this procedure can be adapted to cases where the contamination is more significantly constrained based on physical grounds and give results accordingly. Thus, although being a prospect, we would like to emphasise the information that this method can provide by focusing on the slow glitch trend and directly dealing with the frequencies rather than their combination.
%-------------------------------------------------------------------

\begin{acknowledgements}
    The authors would like to thank the referee for his or her constructive report and willingness to improve and clarify the content of this article.
\end{acknowledgements}

% WARNING
%-------------------------------------------------------------------
% Please note that we have included the references to the file aa.dem in
% order to compile it, but we ask you to:
%
% - use BibTeX with the regular commands:
%   \bibliographystyle{aa} % style aa.bst
%   \bibliography{Yourfile} % your references Yourfile.bib
%
% - join the .bib files when you upload your source files
%-------------------------------------------------------------------
\bibliographystyle{aa}
\hypertarget{paper1}{~}
\bibliography{src.bib}

\begin{thebibliography}{41}
\expandafter\ifx\csname natexlab\endcsname\relax\def\natexlab#1{#1}\fi

\bibitem[{{Aigrain} {et~al.}(2015){Aigrain}, {Hodgkin}, {Irwin}, {Lewis}, \&
  {Roberts}}]{Aigrain2015}
{Aigrain}, S., {Hodgkin}, S.~T., {Irwin}, M.~J., {Lewis}, J.~R., \& {Roberts},
  S.~J. 2015, \mnras, 447, 2880

\bibitem[{{Aigrain} {et~al.}(2016){Aigrain}, {Parviainen}, \&
  {Pope}}]{Aigrain2016}
{Aigrain}, S., {Parviainen}, H., \& {Pope}, B.~J.~S. 2016, \mnras, 459, 2408

\bibitem[{{Angus} {et~al.}(2018){Angus}, {Morton}, {Aigrain}, {Foreman-Mackey},
  \& {Rajpaul}}]{Angus2018}
{Angus}, R., {Morton}, T., {Aigrain}, S., {Foreman-Mackey}, D., \& {Rajpaul},
  V. 2018, \mnras, 474, 2094

\bibitem[{{Antia} \& {Basu}(1994)}]{Antia1994}
{Antia}, H.~M. \& {Basu}, S. 1994, \aaps, 107, 421

\bibitem[{{Baglin} {et~al.}(2006){Baglin}, {Auvergne}, {Barge}, {Deleuil},
  {Catala}, {Michel}, {Weiss}, \& {COROT Team}}]{Baglin2006}
{Baglin}, A., {Auvergne}, M., {Barge}, P., {et~al.} 2006, in ESA Special
  Publication, Vol. 1306, The CoRoT Mission Pre-Launch Status - Stellar
  Seismology and Planet Finding, ed. M.~{Fridlund}, A.~{Baglin}, J.~{Lochard},
  \& L.~{Conroy}, 33

\bibitem[{{Ballot} {et~al.}(2004){Ballot}, {Turck-Chi{\`e}ze}, \&
  {Garc{\'\i}a}}]{Ballot2004}
{Ballot}, J., {Turck-Chi{\`e}ze}, S., \& {Garc{\'\i}a}, R.~A. 2004, \aap, 423,
  1051

\bibitem[{{Bender} \& {Orszag}(1978)}]{Bender1978}
{Bender}, C.~M. \& {Orszag}, S.~A. 1978, {Advanced Mathematical Methods for
  Scientists and Engineers}

\bibitem[{{Broomhall} {et~al.}(2014){Broomhall}, {Miglio}, {Montalb{\'a}n},
  {Eggenberger}, {Chaplin}, {Elsworth}, {Scuflaire}, {Ventura}, \&
  {Verner}}]{Broomhall2014}
{Broomhall}, A.~M., {Miglio}, A., {Montalb{\'a}n}, J., {et~al.} 2014, MNRAS,
  440, 1828

\bibitem[{{Chandrasekhar}(1964)}]{Chandrasekhar1964}
{Chandrasekhar}, S. 1964, A\&A, 139, 664

\bibitem[{{Christensen-Dalsgaard} \& {Perez Hernandez}(1992)}]{JCD1992}
{Christensen-Dalsgaard}, J. \& {Perez Hernandez}, F. 1992, \mnras, 257, 62

\bibitem[{{Christensen-Dalsgaard} \& {Thompson}(1997)}]{JCD1997}
{Christensen-Dalsgaard}, J. \& {Thompson}, M.~J. 1997, \mnras, 284, 527

\bibitem[{{Deheuvels} \& {Michel}(2010)}]{Deheuvels2010}
{Deheuvels}, S. \& {Michel}, E. 2010, Astronomische Nachrichten, 331, 929

\bibitem[{{Dziembowski} {et~al.}(1990){Dziembowski}, {Pamyatnykh}, \&
  {Sienkiewicz}}]{Dziembowski1990}
{Dziembowski}, W.~A., {Pamyatnykh}, A.~A., \& {Sienkiewicz}, R. 1990, MNRAS,
  244, 542

\bibitem[{{Gilliland} {et~al.}(2010){Gilliland}, {Brown},
  {Christensen-Dalsgaard}, {Kjeldsen}, {Aerts}, {Appourchaux}, {Basu},
  {Bedding}, {Chaplin}, {Cunha}, {De Cat}, {De Ridder}, {Guzik}, {Handler},
  {Kawaler}, {Kiss}, {Kolenberg}, {Kurtz}, {Metcalfe}, {Monteiro}, {Szab{\'o}},
  {Arentoft}, {Balona}, {Debosscher}, {Elsworth}, {Quirion}, {Stello},
  {Su{\'a}rez}, {Borucki}, {Jenkins}, {Koch}, {Kondo}, {Latham}, {Rowe}, \&
  {Steffen}}]{Gilliland2010}
{Gilliland}, R.~L., {Brown}, T.~M., {Christensen-Dalsgaard}, J., {et~al.} 2010,
  \pasp, 122, 131

\bibitem[{{Gough}(1990)}]{Gough1990}
{Gough}, D.~O. 1990, {Comments on Helioseismic Inference}, ed. Y.~{Osaki} \&
  H.~{Shibahashi}, Vol. 367, 283

\bibitem[{{Gough}(2002)}]{Gough2002}
{Gough}, D.~O. 2002, in ESA Special Publication, Vol. 485, Stellar Structure
  and Habitable Planet Finding, ed. B.~{Battrick}, F.~{Favata}, I.~W.
  {Roxburgh}, \& D.~{Galadi}, 65--73

\bibitem[{{Gough} \& {Thompson}(1991)}]{Gough1991}
{Gough}, D.~O. \& {Thompson}, M.~J. 1991, {The inversion problem.}, 519--561

\bibitem[{{Houdayer} {et~al.}(2021){Houdayer}, {Reese}, {Goupil}, \&
  {Lebreton}}]{Houdayer2021}
{Houdayer}, P.~S., {Reese}, D.~R., {Goupil}, M.-J., \& {Lebreton}, Y. 2021,
  \aap, 655, A85

\bibitem[{{Houdek} \& {Gough}(2007)}]{Houdek2007}
{Houdek}, G. \& {Gough}, D.~O. 2007, MNRAS, 375, 861

\bibitem[{{Houdek} \& {Gough}(2011)}]{Houdek2011}
{Houdek}, G. \& {Gough}, D.~O. 2011, \mnras, 418, 1217

\bibitem[{{Kjeldsen} {et~al.}(2008){Kjeldsen}, {Bedding}, \&
  {Christensen-Dalsgaard}}]{Kjeldsen2008}
{Kjeldsen}, H., {Bedding}, T.~R., \& {Christensen-Dalsgaard}, J. 2008, \apjl,
  683, L175

\bibitem[{Lopes {et~al.}(1997)Lopes, Turck-Chieze, Michel, \&
  Goupil}]{Lopes1997}
Lopes, I., Turck-Chieze, S., Michel, E., \& Goupil, M.-J. 1997, The
  Astrophysical Journal, 480, 794

\bibitem[{{Lund} {et~al.}(2017){Lund}, {Silva Aguirre}, {Davies}, {Chaplin},
  {Christensen-Dalsgaard}, {Houdek}, {White}, {Bedding}, {Ball}, {Huber},
  {Antia}, {Lebreton}, {Latham}, {Handberg}, {Verma}, {Basu}, {Casagrande},
  {Justesen}, {Kjeldsen}, \& {Mosumgaard}}]{Lund2017}
{Lund}, M.~N., {Silva Aguirre}, V., {Davies}, G.~R., {et~al.} 2017, \apj, 835,
  172

\bibitem[{{Monteiro} {et~al.}(1994){Monteiro}, {Christensen-Dalsgaard}, \&
  {Thompson}}]{Monteiro1994}
{Monteiro}, M.~J.~P.~F.~G., {Christensen-Dalsgaard}, J., \& {Thompson}, M.~J.
  1994, A\&A, 283, 247

\bibitem[{{Monteiro} \& {Thompson}(1998)}]{Monteiro1998}
{Monteiro}, M.~J.~P.~F.~G. \& {Thompson}, M.~J. 1998, in IAU Symposium, Vol.
  185, New Eyes to See Inside the Sun and Stars, ed. F.-L. {Deubner},
  J.~{Christensen-Dalsgaard}, \& D.~{Kurtz}, 317

\bibitem[{{Monteiro} \& {Thompson}(2005)}]{Monteiro2005}
{Monteiro}, M. J.~P.~F.~G. \& {Thompson}, M.~J. 2005, MNRAS, 361, 1187

\bibitem[{{Noll} {et~al.}(2021){Noll}, {Deheuvels}, \& {Ballot}}]{Noll2021}
{Noll}, A., {Deheuvels}, S., \& {Ballot}, J. 2021, \aap, 647, A187

\bibitem[{{Perez Hernandez} \&
  {Christensen-Dalsgaard}(1994)}]{PerezHernandez1994}
{Perez Hernandez}, F. \& {Christensen-Dalsgaard}, J. 1994, \mnras, 269, 475

\bibitem[{{Rasmussen} \& {Williams}(2006)}]{Rasmussen2006}
{Rasmussen}, C.~E. \& {Williams}, C. K.~I. 2006, {Gaussian Processes for
  Machine Learning}

\bibitem[{{Reese} {et~al.}(2016){Reese}, {Buldgen}, \& {Zharkov}}]{Reese2016b}
{Reese}, D., {Buldgen}, G., \& {Zharkov}, S. 2016, {InversionKit: Linear
  inversions from frequency data}

\bibitem[{{Ricker} {et~al.}(2015){Ricker}, {Winn}, {Vanderspek}, {Latham},
  {Bakos}, {Bean}, {Berta-Thompson}, {Brown}, {Buchhave}, {Butler}, {Butler},
  {Chaplin}, {Charbonneau}, {Christensen-Dalsgaard}, {Clampin}, {Deming},
  {Doty}, {De Lee}, {Dressing}, {Dunham}, {Endl}, {Fressin}, {Ge}, {Henning},
  {Holman}, {Howard}, {Ida}, {Jenkins}, {Jernigan}, {Johnson}, {Kaltenegger},
  {Kawai}, {Kjeldsen}, {Laughlin}, {Levine}, {Lin}, {Lissauer}, {MacQueen},
  {Marcy}, {McCullough}, {Morton}, {Narita}, {Paegert}, {Palle}, {Pepe},
  {Pepper}, {Quirrenbach}, {Rinehart}, {Sasselov}, {Sato}, {Seager},
  {Sozzetti}, {Stassun}, {Sullivan}, {Szentgyorgyi}, {Torres}, {Udry}, \&
  {Villasenor}}]{Ricker2015}
{Ricker}, G.~R., {Winn}, J.~N., {Vanderspek}, R., {et~al.} 2015, Journal of
  Astronomical Telescopes, Instruments, and Systems, 1, 014003

\bibitem[{{Roxburgh} \& {Vorontsov}(1994{\natexlab{a}})}]{Roxburgh1994a}
{Roxburgh}, I.~W. \& {Vorontsov}, S.~V. 1994{\natexlab{a}}, \mnras, 268, 143

\bibitem[{{Roxburgh} \& {Vorontsov}(1994{\natexlab{b}})}]{Roxburgh1994b}
{Roxburgh}, I.~W. \& {Vorontsov}, S.~V. 1994{\natexlab{b}}, \mnras, 268, 880

\bibitem[{{Sonoi} {et~al.}(2015){Sonoi}, {Samadi}, {Belkacem}, {Ludwig},
  {Caffau}, \& {Mosser}}]{Sonoi2015}
{Sonoi}, T., {Samadi}, R., {Belkacem}, K., {et~al.} 2015, \aap, 583, A112

\bibitem[{{Stassun} {et~al.}(2019){Stassun}, {Oelkers}, {Paegert}, {Torres},
  {Pepper}, {De Lee}, {Collins}, {Latham}, {Muirhead}, {Chittidi},
  {Rojas-Ayala}, {Fleming}, {Rose}, {Tenenbaum}, {Ting}, {Kane}, {Barclay},
  {Bean}, {Brassuer}, {Charbonneau}, {Ge}, {Lissauer}, {Mann}, {McLean},
  {Mullally}, {Narita}, {Plavchan}, {Ricker}, {Sasselov}, {Seager}, {Sharma},
  {Shiao}, {Sozzetti}, {Stello}, {Vanderspek}, {Wallace}, \&
  {Winn}}]{Stassun2019}
{Stassun}, K.~G., {Oelkers}, R.~J., {Paegert}, M., {et~al.} 2019, \aj, 158, 138

\bibitem[{{Tassoul}(1980)}]{Tassoul1980}
{Tassoul}, M. 1980, ApJS, 43, 469

\bibitem[{{Vandakurov}(1967)}]{Vandakurov1967}
{Vandakurov}, Y.~V. 1967, \azh, 44, 786

\bibitem[{{Verma} {et~al.}(2014){Verma}, {\GG{a}}{Faria}, {Antia}, {Basu},
  {Mazumdar}, {Monteiro}, {Appourchaux}, {Chaplin}, {Garc{\'\i}a}, \&
  {Metcalfe}}]{Verma2014a}
{Verma}, K., {\GG{a}}{Faria}, J.~P., {Antia}, H.~M., {et~al.} 2014, \apj, 790,
  138

\bibitem[{{Verma} {et~al.}(2017){Verma}, {Raodeo}, {Antia}, {Mazumdar}, {Basu},
  {Lund}, \& {Silva Aguirre}}]{Verma2017}
{Verma}, K., {Raodeo}, K., {Antia}, H.~M., {et~al.} 2017, \apj, 837, 47

\bibitem[{{Verma} {et~al.}(2019){Verma}, {Raodeo}, {Basu}, {Silva Aguirre},
  {Mazumdar}, {Mosumgaard}, {Lund}, \& {Ranadive}}]{Verma2019}
{Verma}, K., {Raodeo}, K., {Basu}, S., {et~al.} 2019, \mnras, 483, 4678

\bibitem[{{Vorontsov} {et~al.}(1992){Vorontsov}, {Baturin}, \&
  {Pamiatnykh}}]{Vorontsov1992}
{Vorontsov}, S.~V., {Baturin}, V.~A., \& {Pamiatnykh}, A.~A. 1992, \mnras, 257,
  32

\end{thebibliography}

% APPENDIX
%-------------------------------------------------------------------
\appendix
\section{Details about the recovery of the posterior distributions \label{AppendixI}}

We provide here details about the Bayesian inference on which Sections~\ref{FIT} and \ref{DISCUSSION} rely.

\subsection{Fitting the frequencies using $\delta\vec{\upsigma^2}_\mathrm{mod}(\delta\vec{\uptheta})$ \label{AI1}}

We will first provide a prior distribution on $\vec{\uptheta}$, by decomposing it on each parameter (assumed to be a priori independent):
\begin{equation}
    \label{eq:prior}
    p(\vec{\uptheta}) = \prod_i p(\theta_i) =  p(Y_s)p(\psi_\mathrm{CZ})p(\varepsilon)
\end{equation} with $p(X)$ designating the probability density function (PDF) of the random variable $X$. Specifically, we use the following distribution laws 
\begin{align*}
    Y_s &\sim \mathrm{Beta}(\mu=0.25, s=0.1), \\
    -\psi_\mathrm{CZ} &\sim \mathrm{BetaPrime}(\mu=5, s=2), \\
    \varepsilon &\sim \mathrm{BetaPrime}(\mu=0.02, s=0.01).
\end{align*}
The natural parameters $(\alpha, \beta)$ of these two laws can easily be determined from our parameterisation via the mean, $\mu$, and standard deviation, $s$:
\begin{align*}
    \mathrm{Beta:~~~~~~~~~~~~} \alpha &= \mu\left(\mathrm{v}_- - 1\right), \quad \beta = (1-\mu)\left(\mathrm{v}_- - 1\right)\\
    \mathrm{BetaPrime: ~~} \alpha &= \mu\left(\mathrm{v}_+ + 1\right), \quad \beta = \mathrm{v}_+ + 2
\end{align*} with $\mathrm{v}_\pm = \mu(1\pm\mu)/s^2$.
We chose the above distributions because they are defined on appropriate domains and have advantageous properties: $0\leq Y_s\leq 1$, $\psi_\mathrm{CZ} < 0$ (the convective region in solar-like pulsators cannot be degenerate), $\varepsilon > 0$ (by design). The means given correspond more or less to solar values but with a sufficient variability to remain rather uninformative. The prior distribution on $\delta\vec{\uptheta}$ trivially follows from:
\begin{equation}
    \delta\vec{\uptheta} = \vec{\uptheta} - \vec{\uptheta}^\odot
\end{equation} with $\vec{\uptheta}^\odot = [0.255, -5, 0.015]^\mathsf{T}$ our (fixed) reference. We will assume that the observed glitch distribution differs from the model only by an unbiased multivariate Gaussian, that is:
\begin{equation}
    \delta\vec{\upsigma^2}|\delta\vec{\uptheta} = \delta\vec{\upsigma^2}_\mathrm{mod}(\delta\vec{\uptheta}) + \vec{\mathrm{n}} \mbox{ ~with~ } \vec{\mathrm{n}} \sim \mathcal{N}\left(\vec{\mathrm{0}},\Sigma_{\delta\sigma^2}\right) 
\end{equation}with  $\delta\vec{\upsigma^2}_\mathrm{mod}(\delta\vec{\uptheta})$ denoting the vector composed of all modelled (quadratic) frequency deviations from the reference given by Eq.~\eqref{eq:glitch_mod_lin}. Our modelling of the noise $\vec{\mathrm{n}}$ remains very rudimentary: we have assumed a frequency-independent and uncorrelated error distribution on $\delta\sigma$, i.e. $\Sigma_{\delta\sigma} = {s_{\vec{\mathrm{n}}}}^2\mathrm{I}$. From there, the noise distribution is easily propagated from $\delta\vec{\upsigma}$ to $\delta\vec{\upsigma^2}$:
\begin{equation}
    \label{eq:Sigma_n}
    \Sigma_{\delta\sigma^2} = (2\mathrm{diag}(\vec{\upsigma}))({s_{\vec{\mathrm{n}}}}^2\mathrm{I})(2\mathrm{diag}(\vec{\upsigma}))^\mathsf{T} = (2{s_{\vec{\mathrm{n}}}}\mathrm{diag}(\vec{\upsigma}))^2
\end{equation} with $\mathrm{diag}(\vec{\upsigma})$ the diagonal matrix composed of the $\sigma$. From Bayes' theorem, the most plausible values of $\delta\vec{\uptheta}|\delta\vec{\upsigma^2}$ minimise the following function (explicitly written in terms of the distributions introduced above):
\begin{equation}
    -\ln p(\delta\vec{\uptheta}|\delta\vec{\upsigma^2}) = -\ln p(\delta\vec{\uptheta}) - \ln p( \delta\vec{\upsigma^2}|\delta\vec{\uptheta})+\ln p(\delta\vec{\upsigma^2})
\end{equation} with 
\begin{equation}
    \label{eq:marginal_likelihood}
    p(\delta\vec{\upsigma^2}) = \int p(\delta\vec{\uptheta})p( \delta\vec{\upsigma^2}|\delta\vec{\uptheta})\,d\delta\vec{\uptheta},
\end{equation}the marginal likelihood that can be treated as a normalising constant in this case. 

\subsection{Fitting the frequencies using $\delta \vec{\mathrm{y}}_\mathrm{mod}(\delta\vec{\uptheta})$ \label{AI2}}

In the presence of a contamination by surface effects, the observed frequency differences will be broken down into two components (in addition to the noise $\vec{\mathrm{n}}$):
\begin{equation}
    \label{eq:dnu2_tot}
    \delta\vec{\upsigma^2} = \delta \vec{\upsigma^2}_\mathrm{ion} + \delta \vec{\upsigma^2}_\mathrm{surf} + \vec{\mathrm{n}}
\end{equation} where $\delta \vec{\upsigma^2}_\mathrm{ion}$ designates the ionisation glitch and $\delta \vec{\upsigma^2}_\mathrm{surf}$ the deviation caused by surface effects. The method presented in Section~\ref{diagnostic} consists in introducing the combination:
\begin{equation}
    \delta\vec{\mathrm{y}} = \mathrm{C}\delta\vec{\upsigma^2} = \mathrm{C}\delta \vec{\upsigma^2}_\mathrm{ion} + \mathrm{C}\delta \vec{\upsigma^2}_\mathrm{surf} + \mathrm{C}\vec{\mathrm{n}}
\end{equation} in the hope that $\mathrm{C}\delta \vec{\upsigma^2}_\mathrm{surf} \ll \mathrm{C}\delta \vec{\upsigma^2}_\mathrm{ion}$. In such a case, it is possible to approximate $\delta\vec{\mathrm{y}}$ by the sole ionisation diagnostic model $\mathrm{C}\delta \vec{\upsigma^2}_\mathrm{ion}(\delta\vec{\uptheta})$  (given by Eq.~\eqref{eq:diff_glitch_mod_lin}) together with $\mathrm{C}\vec{\mathrm{n}}$. In the case considered in the main part where $\mathrm{C} = \mathrm{D}^2$, the noise covariance matrix is given by:
\begin{equation}
    \label{eq:cov_diagnostic}
    \Sigma_{\delta\mathrm{y}} = \mathrm{D}^2(2{s_{\vec{\mathrm{n}}}}\mathrm{diag}(\vec{\upsigma}))^2(\mathrm{D}^2)^\mathsf{T}.
\end{equation}

\subsection{Fitting the frequencies using $\delta\vec{\upsigma^2}_\mathrm{mod}(\delta\vec{\uptheta}, \vgp)$ \label{AI3}}

A comprehensive introduction to the theory of GPs is given in \cite{Rasmussen2006} and most of the notations appearing in this appendix are based on it. For instance, we will write:
\begin{equation}
    \label{eq:def_gp}
    \delta\sigma^2_\mathrm{surf}(\sigma) \sim \mathcal{GP}\left(0,k_\mathrm{surf}(\sigma, \sigma')\right)
\end{equation}to designate a model of the surface effects following the GP with mean function $m_\mathrm{surf}(\sigma) = 0$ and covariance function $k_\mathrm{surf}(\sigma, \sigma')$. Although it may seem surprising, the choice of a zero mean a priori is rather common and enables us to avoid expressing a parametric form of $m_\mathrm{surf}(\sigma)$ explicitly. Constrained by the observations $\mathcal{D} = \left\{\vec{\upsigma}, \delta \vec{\upsigma^2}\right\}$, it is clear that this mean will no longer be zero but will instead be expressed in terms of $k_\mathrm{surf}(~\cdot~, \vec{\upsigma})$ and $\delta \vec{\upsigma^2}$. The explicit definition of the process in Eq.~\eqref{eq:def_gp} then boils down to expressing the covariance function, $k_\mathrm{surf}(\sigma, \sigma')$, which gives the covariance between any choice of two frequencies $(\sigma, \sigma')$. Many astrophysical applications of GPs such as corrections of systematics and detrending of K2 light curves or estimates of stellar rotation periods \citep[e.g.][]{Aigrain2015,Aigrain2016,Angus2018} already make use of the ``squared exponential'' (SE) and ``quasi-periodic'' (QP) covariance functions. In this paper, in order to satisfy the constraints described above, we will choose $k_\mathrm{surf}$ to be the product of a sixth order ``homogeneous polynomial'' (HP) and a SE covariance function:
\begin{equation}
    \label{eq:ksurf}
    \begin{split}
        k_\mathrm{surf}(\sigma, \sigma') &= k_\mathrm{HP}(\sigma, \sigma')\times k_\mathrm{SE}(\sigma, \sigma') \\
        &= \frac{\left(\sigma\sigma'\right)^6}{\beta^2} \times \exp\left(-\frac{ M(\sigma-\sigma')^2}{2}\right).
    \end{split}
\end{equation}
In this function, each part acts as follows:
\begin{itemize}
    \item[$\bullet$] The HP covariance function reduces the variance of the frequency, $\sigma$, when its value diminishes. Thus the process will take on values close to its mean (in this case $0$) at low frequencies and allow a larger dispersion at higher frequencies. The ``envelope'' of the process (in the sense of equally likely values as a function of frequency) is polynomial and varies as $\sigma^6/\beta$. The choice of the exponent ($6$) is due to the observation that the surface effects vary approximately as $\delta\sigma_\textrm{surf}\propto\sigma^5$ in the solar case \citep{Kjeldsen2008}, thus making $\delta\sigma^2_\textrm{surf}\propto\sigma^6$. This choice is somewhat arbitrary although we will see that the process can widely deviate from this envelope.
    \item[$\bullet$] The SE covariance function reduces the covariance between two frequencies, $\sigma$ and $\sigma'$, that are distant in value. This introduces a characteristic scale of variation that behaves as $\sqrt{1/M}$ and allows for a progressive departure from a strict polynomial relationship. This covariance function therefore adds flexibility to the process.
\end{itemize}

The process $\delta\sigma^2_\mathrm{surf}(\sigma)$ can then be seen as parameterised by the two values $\beta^2$ and $M$ (strictly speaking, they are hyperparameters), which influence the probability distribution of finding the value $\delta\sigma^2_\textrm{surf}(\sigma)$ at frequency $\sigma$. Both of these parameters being positive (and potentially very large), we will refer to this parameterisation as $\vgp = [\log \beta^2, \log M]^\mathsf{T}$ (and to the corresponding GP as $\delta\sigma^2_{\mathrm{surf},\vgp}(\sigma)$). Based on this GP, our parameterised model of surface effects in Eq.~\eqref{eq:df2_ion_surf} directly follows from the estimate of the mean process on $\vec{\upsigma}$ for a fixed value of $\vgp$:
\begin{equation}
    \delta \vec{\upsigma^2}_\mathrm{surf}(\vgp) = \overline{\delta\sigma^2}_{\mathrm{surf},\vgp}(\vec{\upsigma}).
\end{equation}

We will also adopt the notation:
\begin{equation}
    \mathrm{K}_\mathrm{surf}(\vgp) = k_{\mathrm{surf},\vgp}(\vec{\upsigma}, \vec{\upsigma})
\end{equation} to designate the covariance matrix resulting from applying $k_{\mathrm{surf},\vgp}$ to any pair of frequencies in $\vec{\upsigma}$.

Given an observation of surface effects, it would then be fairly easy to deduce from the GP prior (given in Eq.~\eqref{eq:def_gp}) the process a posteriori. Such a case never happens in practice, however. As highlighted by Eq.~\eqref{eq:dnu2_tot}, the observation of a surface effect will never occur independently of that of the glitch and of a stochastic component. Even if the ionisation glitch component is purely deterministic (unlike the other two), it is useful to write the entire sum as a GP itself to exploit its properties. Thus, the process representing the entirety of the observed frequency shift can be written as:
\begin{equation}
    \label{eq:def_gp_tot}
    \delta\sigma^2(\sigma) \sim \mathcal{GP}\left(\delta\sigma^2_{\mathrm{ion},\delta\vec{\uptheta}}(\sigma), k_{\mathrm{surf},\vgp}(\sigma, \sigma')+k_n(\sigma, \sigma')\right)
\end{equation} with $\delta\sigma^2_{\mathrm{ion},\delta\vec{\uptheta}}(\sigma)$ the prediction of the model given by Eq.~\eqref{eq:glitch_mod_lin} at~$\sigma$, $k_n(\sigma,\sigma')$ the (non-stationary) white noise covariance function defined as
\begin{equation}
    \label{eq:cov_fun_n}
    k_n(\sigma, \sigma') = (2{s_{\vec{\mathrm{n}}}}\sigma)^2\delta(\sigma'-\sigma)
\end{equation}
and associated with the covariance matrix $\Sigma_{\delta\sigma^2}$, $\delta(x)$ being the Dirac function. Indeed, it is clear from Eqs.~\eqref{eq:Sigma_n} and \eqref{eq:cov_fun_n} that we have $k_n(\vec{\upsigma}, \vec{\upsigma}) = \Sigma_{\delta\sigma^2}$. We note that the deterministic aspect of the ionisation glitch component results in the addition of a mean value to the process only without any additional covariance compared to Eq.~\eqref{eq:def_gp}. Also, as for $\vgp$, $\delta\vec{\uptheta}$ can be considered here as a hyperparameter in the sense that it designates a parameter of the underlying process (its mean function) and not of a particular realisation $\delta \vec{\upsigma^2}$.

By noticing that $\delta\vec{\upsigma^2}|\delta\vec{\uptheta},\vgp$ simply follows the normal law $\mathcal{N}(\delta \vec{\upsigma^2}_\mathrm{ion}(\delta\vec{\uptheta}), \mathrm{K}_\mathrm{surf}(\vgp)+\Sigma_{\delta\sigma^2})$ from Eq.~\eqref{eq:def_gp_tot}, the resulting marginalised likelihood is given by:
\begin{equation}
    \label{eq:marginal_likelihood2}
    \begin{split}
        &\ln p(\delta\vec{\upsigma^2}|\delta\vec{\uptheta},\vgp) = \\
        &-\frac{1}{2}\left(\delta\vec{\upsigma^2} - \delta \vec{\upsigma^2}_\mathrm{ion}(\delta\vec{\uptheta})\right)^\mathsf{T}{\left[\mathrm{K}_\mathrm{surf}(\vgp)+\Sigma_{\delta\sigma^2}\right]}^{-1}\left(\delta\vec{\upsigma^2}- \delta \vec{\upsigma^2}_\mathrm{ion}(\delta\vec{\uptheta})\right) \\
        &-\frac{1}{2}\ln\left|\mathrm{K}_\mathrm{surf}(\vgp)+\Sigma_{\delta\sigma^2}\right|
        -\frac{N}{2}\ln 2\pi
    \end{split}
\end{equation} with $N$ denoting the number of radial orders considered. Here, the designation `marginal' may seem obscure when compared to Eq.~\eqref{eq:marginal_likelihood}, where it referred to an integration over any possible $\delta\vec{\uptheta}$. In Eq.~\eqref{eq:marginal_likelihood2}, we refer instead to a marginalisation over the realisations of the GP for a fixed value of $\delta\vec{\uptheta}$ and $\vgp$ (hence their designation by the term hyperparameters). We note that this expression is analytically tractable only because we have assumed our frequency shift model to be a GP. Subsequently, accounting for the prior distributions, we will sample the following hyperparameter posterior:
\begin{equation}
    \begin{split}
        \ln p(\delta\vec{\uptheta},\vgp|\delta\vec{\upsigma^2}) &= \ln p(\delta\vec{\uptheta},\vgp) + \ln p( \delta\vec{\upsigma^2}|\delta\vec{\uptheta},\vgp) - \ln p(\delta\vec{\upsigma^2})\\
        &= \ln p(\delta\vec{\uptheta}) + \ln p(\vgp) + \ln p( \delta\vec{\upsigma^2}|\delta\vec{\uptheta},\vgp) - \ln p(\delta\vec{\upsigma^2})
    \end{split}
\end{equation} with $p(\delta\vec{\uptheta})$ resulting from Eq.~\eqref{eq:prior}, $p(\vgp)$ being taken as an improper (in the sense of Bayesian inference) uniform prior, $p(\delta\vec{\upsigma^2}|\delta\vec{\uptheta},\vgp)$ being given by Eq.~\eqref{eq:marginal_likelihood2} and 
\begin{equation}
    p(\delta\vec{\upsigma^2}) = \iint p(\delta\vec{\uptheta})p(\vgp)p( \delta\vec{\upsigma^2}|\delta\vec{\uptheta},\vgp)\,d\delta\vec{\uptheta}d\vgp
\end{equation} being the normalising constant.

\section{Parameter recovery with additional frequencies \label{AppendixII}}

\begin{figure}
   \centering
   \includegraphics[width=9cm]{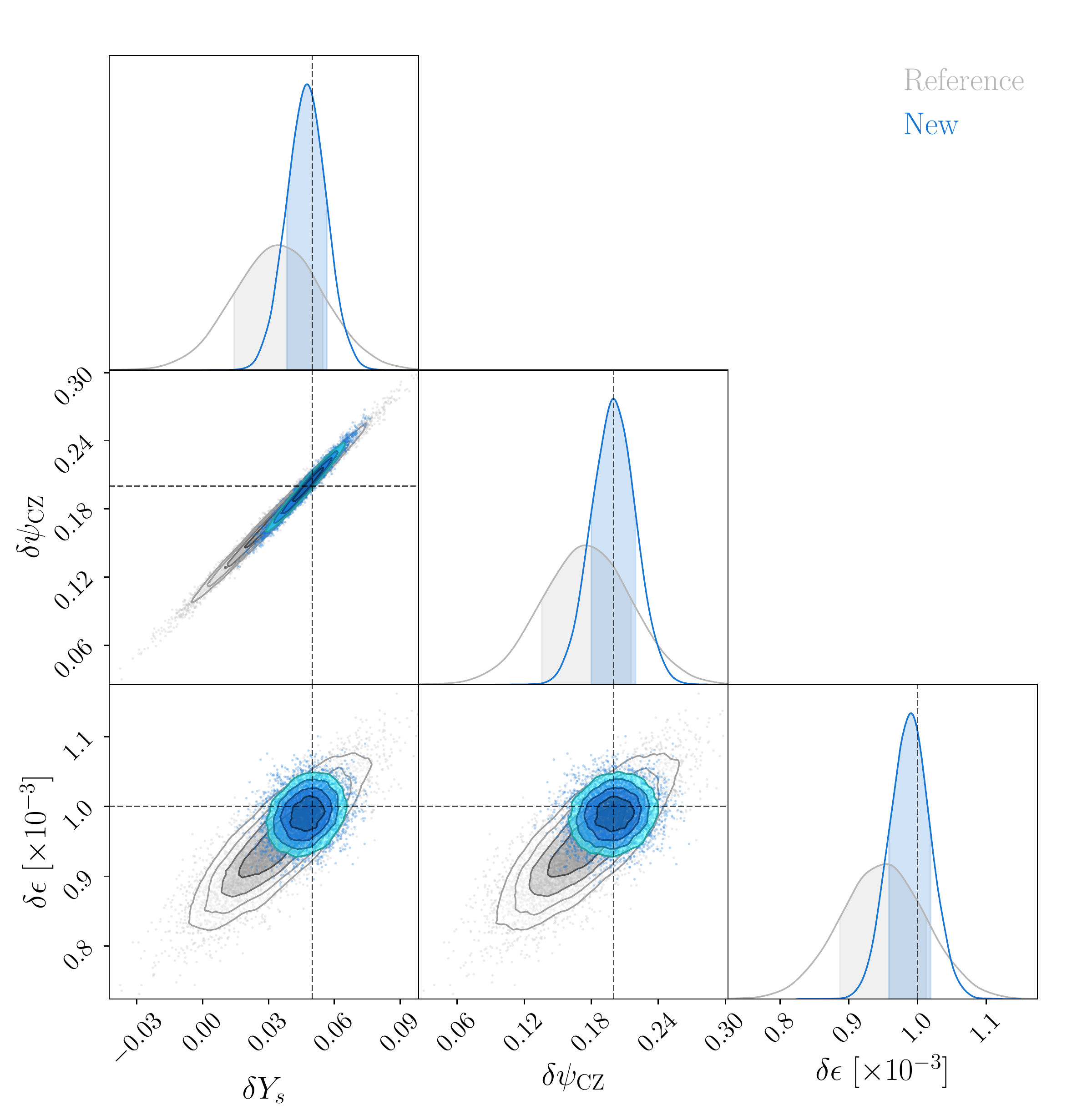}
   \caption{MCMC estimate of the distribution $p(\delta\vec{\uptheta}|\delta\vec{\upsigma^2})$ with the additional low order frequencies $7\leq n \leq 10$ (in blue) in comparison with the reference shown in Fig.~\ref{mcmc_f.pdf} (in grey). The dashed lines indicate the value of the true $\delta\vec{\uptheta}_\mathrm{true} = [0.05, 0.2, 1\times10^{-3}]^\mathsf{T}$ used to generate $\delta\sigma^2$ (the PDF gives at this point: $\ln p(\delta\vec{\uptheta}_\mathrm{true}|\delta\vec{\upsigma^2}) = 16.91$). The mode of this distribution is $\delta\vec{\uptheta}_\mathrm{MAP} = [0.047,0.200,0.989\times10^{-3}]^\mathsf{T}$.}
        \label{mcmc_f_plus.pdf}%
    \end{figure}

\begin{figure}
   \centering
   \includegraphics[width=9cm]{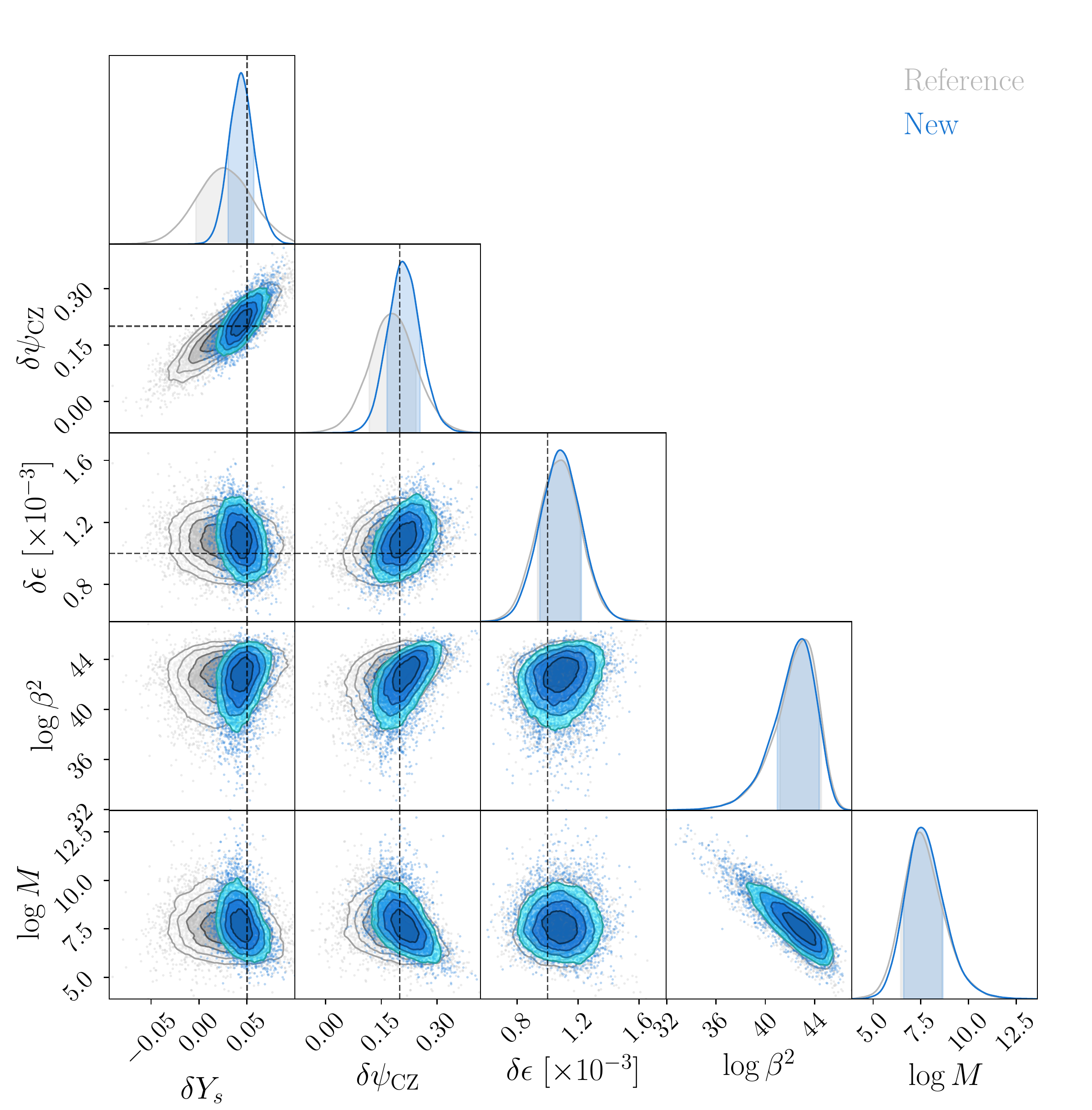}
   \caption{MCMC estimate of the distribution $p(\delta\vec{\uptheta}, \vgp|\delta\vec{\upsigma^2})$ with the additional low order frequencies $7\leq n \leq 10$ (in blue) in comparison with the reference shown in Fig.~\ref{mcmc_gp.pdf} (in grey). The dashed lines indicate the value of $\delta\vec{\uptheta}_\mathrm{true} = [0.05, 0.2, 1\times10^{-3}]^\mathsf{T}$ used to generate $\delta \vec{\upsigma^2}$ (the marginalised PDF over $\vgp$ gives at this point: $\ln p(\delta\vec{\uptheta}_\mathrm{true}|\delta\vec{\upsigma^2}) = 14.37$). The mode of this distribution is given by $\delta\vec{\uptheta}_\mathrm{MAP} = [0.046,0.214,1.083\times10^{-3}]^\mathsf{T}$ and $\vgp_\textrm{MAP} = [43.2, 7.57]^\mathsf{T}$.}
        \label{mcmc_gp_plus.pdf}%
    \end{figure}

Considering all the distribution sampling performed so far, it may be interesting to try to quantify the impact of the various nuisances on the parameter recovery. In our case, besides potential imperfections of the model we presented, a clearly identifiable source of error lies in the in the absence of very low degree frequencies. Indeed, these frequencies contain the most information on the slowly varying component of the glitch, which we expect to be essential for a complete retrieval of the parameters. To this end, we present in Figs.~\ref{mcmc_f_plus.pdf} and \ref{mcmc_gp_plus.pdf} the sampling of the distributions shown in Figs.~\ref{mcmc_f.pdf} and \ref{mcmc_gp.pdf} with additional frequencies corresponding to $7\leq n \leq 10$ (the noise level on the oscillation frequencies, ${s_{\vec{\mathrm{n}}}}$, remains the same).

Figure~\ref{mcmc_f_plus.pdf} thus shows the significant improvement in the parameter recovery when fitting the frequencies $\delta\vec{\upsigma^2}$ in the ideal case (to be compared with the distribution found in Fig.~\ref{mcmc_f.pdf}, shown in grey). The mode, $\delta\vec{\uptheta}_\mathrm{MAP} = [0.047,0.200,0.989\times10^{-3}]^\mathsf{T}$, is now much closer to the true value $\delta\vec{\uptheta}_\mathrm{true}$ and the extent of the distribution in general has considerably narrowed, now giving $\ln p(\delta\vec{\uptheta}_\mathrm{true}|\delta\vec{\upsigma^2}) = 16.91$. We will also note the clearly distinct nature of the $\delta Y_s-\delta\psi_\mathrm{CZ}$ degeneracy with its physical origin and the correlation of these two parameters with $\delta\varepsilon$. The first one, although shortened, conserves the same shape while the second one completely disappears to give fully uncorrelated parameters. Thus, if the latter is indeed the result of a lack of information on the slow component associated with hydrogen (cf. Sections~\ref{FIT} and \ref{DISCUSSION}), we witness here the fundamentally different nature of the $\delta Y_s-\delta\psi_\mathrm{CZ}$ degeneracy. If its extent can be reduced as is done here (a decrease in the noise level ${s_{\vec{\mathrm{n}}}}$ also diminishes its size), any glitch estimate of the helium abundance will be confronted with this elongated shape, which results from an ambivalence within the star's structure itself.

Figure~\ref{mcmc_gp_plus.pdf} shows a similar test for the case of a contamination by surface effects. We therefore choose as a reference the distribution $p(\delta\vec{\uptheta}, \vgp|\delta\vec{\upsigma^2})$, shown in Fig.~\ref{mcmc_gp.pdf} and appearing here in grey. Interestingly, the additional frequencies did not have any impact on the distribution marginalised over $\delta Y_s$ and $\delta\psi_\mathrm{CZ}$ (6 rightmost panels) where the new distribution perfectly overlaps with the reference. This new information only effectively constrains the distribution on $\delta Y_s$ and $\delta\psi_{\mathrm{CZ}}$ (visible in the uppermost bivariate distribution) and the latter is largely improved. The MAP estimate, $\delta\vec{\uptheta}_\mathrm{MAP} = [0.046,0.214,1.083\times10^{-3}]^\mathsf{T}$, has moved closer to $\delta\vec{\uptheta}_\mathrm{true}$ and the distribution extent has narrowed, resulting in a value of $\ln p(\delta\vec{\uptheta}_\mathrm{true}|\delta\vec{\upsigma^2}) = 14.37$.

We thus emphasis the importance of these very low radial orders, in particular to constrain the physical quantities, especially the helium abundance and the electronic degeneracy parameter. Let us note that adding, on the other hand, high radial orders ($>27$) hardly changes the retrieved distributions.

% MCMC F PLUS FIGURE --------------------------------------
% --------------------------------------------------------

% MCMC GP PLUS FIGURE --------------------------------------
% --------------------------------------------------------
%-------------------------------------------------------------------
\end{document}